\documentclass[preprint,12pt]{elsarticle}

\usepackage{lineno}
\usepackage{makecell}

\usepackage{verbatim}
\usepackage{multirow}

\modulolinenumbers[5]

\journal{}

\usepackage{color}
\usepackage[normalem]{ulem}
\newcommand{\rev}[1]{#1}
\newcommand{\changed}[1]{#1}

\usepackage{amsfonts,amsmath,amssymb,amsthm,amscd}
\usepackage{siunitx}
\usepackage[colorlinks=true,linkcolor=blue,citecolor=blue]{hyperref}

\usepackage{algorithm}
\usepackage[noend]{algpseudocode}

\newcommand{\rd}{\mathrm{d}}
\newcommand{\e}{\mathrm{e}}
\newcommand{\energy}{\varepsilon}
\newcommand{\tstar}{t^\ast}
\DeclareMathOperator{\cdf}{cdf}

\begin{document}

\begin{frontmatter}

\title{Optimized event generator for strong-field QED simulations within the hi-$\chi$ framework}
\author[UNN]{V.~Volokitin}
\author[GU]{J.~Magnusson}
\author[IAP]{A.~Bashinov}
\author[IAP]{E.~Efimenko}
\author[IAP]{A.~Muraviev}
\author[UNN]{I.~Meyerov}

\address[UNN]{Lobachevsky State University of Nizhni Novgorod, Nizhny Novgorod 603950, Russia}
\address[GU]{Department of Physics, University of Gothenburg, SE-41296 Gothenburg, Sweden}
\address[IAP]{Institute of Applied Physics, Russian Academy of Sciences, Nizhny Novgorod 603950, Russia}

\date{\today}

\begin{abstract}

Probabilistic generation of photons and electron-positron pairs due to the processes of strong-field quantum electrodynamics (SFQED) is often the most resource-intensive part of the kinetic simulations required in order to model current and future experimental studies at high-intensity laser facilities. To reduce its computational demands one can exploit tabulation of the precomputed rates, time-step sub-cycling, dynamic down-sampling of particle/photon ensembles and other approaches. As the culmination of previous improvements, the method described here provides the opportunity to make the minimal possible number of rate computations per QED event\changed{. In some of our tests, this method was shown to increase performance by more than an order of magnitude}. The computational routine is publicly available as a part of the open-source framework hi-$\chi$ designed as a Python-controlled toolbox for collaborative development.

\end{abstract}


\end{frontmatter}

\section{Introduction}

Modern high-intensity lasers provide the opportunity to reach electromagnetic field intensities sufficient for invoking processes of strong-field quantum electrodynamics (SFQED), namely non-linear Compton scattering and Breit-Wheeler pair production \cite{dipiazza.rmp.2012}. These processes trigger a variety of phenomena, for both single-particle and collective dynamics of emerging plasmas, with potential applications as novel particle and radiation sources, as well as for fundamental studies relevant to SFQED, plasma physics and astrophysics \cite{gonoskov.arxiv.2021}. 

Numerical simulations with probabilistic account for SFQED processes play an important role in both theoretical studies and experiment design for the upcoming experimental programs in this field. A particularly fruitful methodology is based on extending the Particle-In-Cell (PIC) method with a module that accounts for the SFQED processes via probabilistic generation of macro-photons, electrons and positrons that sample the 3D3P distribution by an ensemble of reasonable size \cite{nerush.prl.2011, elkina.prstab.2011, sokolov.pop.2011, ridgers.jcp.2014, gonoskov.pre.2015, chang.2017, derouillat.cpc.2018, fedeli.arxiv.2021}. The implementations of this approach, also referred to as QED-PIC codes, include EPOCH \cite{ridgers.jcp.2014}, PIConGPU \cite{bussmann.2013}, VPIC 2.0 \cite{bird.2022}, OSIRIS \cite{fonseca.lncs.2002}, Smilei \cite{derouillat.cpc.2018}, Tristan-MP \cite{spitkovsky.2005}, Calder \cite{lifschitz.jcp.2009}, PICADOR \cite{bastrakov.jcs.2012}, PICSAR-QED \cite{fedeli.arxiv.2021} to name a few.

Although the QED-PIC method is practical, it can become computationally demanding due to two main reasons: (1) the prolific generation of new particles/photons and (2) the extremely short time step sometimes required in order to resolve the characteristic temporal interval between SFQED events (as compared to the laser wave period and other macroscopic scales of the simulated processes). The former is commonly addressed by resampling the ensemble through either merging/coalescing macro-particles/ photons \cite{lapenta.jcp.1994, rjasanow.jcp.1996, luu.cpc.2015, vranic.cpc.2015} or by removing some of the macro-particles/photons and redistributing their statistical weight among the remaining ones \cite{timokhin.mnras.2010, nerush.prl.2011, muraviev.cpc.2021, gonoskov.cpc.2022}. 

The way to handle computational demands caused by the short characteristic time between SFQED events depends on the method used for sampling these events. The use of \emph{rejection sampling} implies that for each time step and for each photon/particle two random numbers are generated: one that defines the proposed energy of photon/particle to be generated and another that determines whether the proposal is accepted or rejected \cite{nerush.prl.2011}. This method is particularly simple to implement as it requires the computation of \changed{two special functions, more precisely the first and second synchrotron functions~\cite{gonoskov.pre.2015}, found in some standard libraries}. \changed{The low-energy singularity in the probability distribution for Compton scattering can be resolved  via increased sampling of low energy proposals through the method of \emph{importance sampling} \cite{tokdar2010importance}.}
By setting the requirements on the time step and implementation of time step sub-cycling to meet said requirements, it also becomes possible to meet the general rejection sampling requirements for any simulation setting \cite{gonoskov.pre.2015}. For each particle/photon this procedure permits the generation of arbitrarily many events (even local cascades) during each global time step, making the whole computational method versatile for studies of extreme processes \cite{gonoskov.prx.2017, efimenko.scirep.2018, efimenko.pre.2019}.

However, the use of rejection sampling has a disadvantage: even if the restrictions on the sub-step duration are mildened, the method requires multiple\changed{, and sometimes expensive,} computations of the involved special functions per \changed{accepted} SFQED event. \cite{volokitin.jpcs.2020} The reason for this is that events need to be rare (per time step) in order to be accurately counted. In the limit of the time step being large there is a unit probability of accepting proposals, resulting in one event per time step. This leads to an artificially uniform generation of events and the exclusion of cases when more than a single event happens within a single time sub-step. 

In order to avoid unnecessary computations of the special functions one can use \changed{\emph{inverse sampling}, in which a distribution is sampled through its cumulative distribution, which} has already been used in the area of PIC simulations to account for ionization processes (see e.g. \cite{nuter.pop.2011}). For SFQED processes the use of inverse sampling implies the following procedure \cite{duclous.ppcf.2011, ridgers.jcp.2014, tamburini.arxiv.2023}. For each particle/photon we generate a uniformly distributed in the unit interval random number $p$ and compute the so-called \textit{optical depth} $d$ by inverting $p = 1 – e^{-d}$. If we represent the probability of an event \emph{not} happening within the time interval $(t, t + t_1)$ as $P(t, t + t_1) = 1 - \exp\left(-\int_t^{t+ t_1} R(\tau)d\tau\right)$, where $R(\tau)$ is the instantaneous event rate over the particle's trajectory, then the exponent plays the role of a cumulative optical depth. As the particle propagates, the integral in the exponent is computed using first-order Eulerian integration until it reaches the value of $d$, at which point the event is assigned to occur. The energy of the outgoing particles is determined randomly using inverse sampling through pre-computed look-up tables that store the inverse cumulative probability as a two-dimensional function of energy and the quantum nonlinearity parameter $\chi$. A new value of optical depth is then generated in order to determine the moment for the next event to happen, and so on. 

In the inverse sampling procedure that was just described one of the most computationally demanding parts is the evaluation of the rate, which is performed once per particle per global time step. This is done in many state-of-the-art QED-PIC codes, including \cite{ridgers.jcp.2014, fedeli.arxiv.2021}. For many cases where the global time step is set by the requirements of the field solver, this is not a very limiting restriction, as it becomes necessary to reevaluate the rate over this time scale anyways in order to account for the field evolution. However, as was the case for the rejection sampling scheme described earlier, the global time step has to be sufficiently small in order to ensure that events are rare. This means that several evaluations of the rate are necessary for the same reason as described before. Furthermore, choosing the global time step according to the highest rate expected to be observed in a simulation can result in additional computational demands.

\changed{While there are many techniques for reducing computational demands, it is difficult to assess performance in an exact and general manner, as performance strongly depends on factors such as problem, method, hardware and various implementation details. We have here described a number of methods that are commonly employed to improve the performance of SFQED computations. However, depending on factors such as those just listed, the performance for any given scenario might naturally vary. Therefore, the development and implementation of a unified methodology for comparing the performance of PIC codes is a very difficult problem. However, we can focus on minimizing the time step sub-cycling. Regardless of the hardware and software environment and implementation features of the various components of the QED-PIC method, reducing the time-step sub-cycling should speed up simulations. That is what this paper focuses on.}

In this article we present the implementation of a method that combines the inverse sampling methodology \changed{described above} with sub-cycling. In doing so, within each global time step we allow multiple SFQED events originated from a single particle/photon\changed{, including from secondary particles}. After the optical depth is reached, we generate new particles, assign optical depths for each of them and continue the process until the end of the global time step. At this point we re-evaluate the rate for all particles/photons in the ensemble, as the global time step often defines the time scale over which the field can no longer be assumed to remain unchanged. For this reason this procedure implies the minimum number of possible rate evaluations: either one per event or one per global time step, depending on whether the optical depth is shorter or longer than the global time step. Furthermore, our implementation\changed{, described in more detail in Section~\ref{sec:method},} also employs high-precision approximations of the cumulative distribution functions with analytical treatment of limiting cases. \changed{The computational accuracy is presented in Section~\ref{sec:validation} and we perform comprehensive testing in Section~\ref{sec:verification}.}

\section{Motivation}
The QED-PIC event generator can be implemented in several different ways, but one thing in common between all of them is that they sample the events in accordance with the partial rates as described later in equations \ref{eq:photon-partial-rate} and \ref{eq:pair-partial-rate}. The two most common ways of sampling the spectra is to employ either rejection sampling or inverse transform sampling. In the former case, an energy value is first picked at random after which rejection sampling is performed on the partial rate $(\rd W/\rd E)$ to determine if the event should occur. In the latter case, it is first decided if the event occurs at all using the event rate ($W = \int (\rd W/\rd E) \,\rd E$), then the energy is determined through inverse transform sampling of the partial rate.

There are advantages and disadvantages to each of the two approaches. For example, the latter requires tabulation of the inverse cumulative distribution function, but in both cases the probability of a QED event to occur within one time step must, naturally, be less than unity. More rigorously, the probability of a single event within one time step must in fact be much smaller than unity in order for the number of events over an extended duration of time to be accurate. The reason for this is that we traditionally neglect the occurrence of multiple events within a single time step. \changed{The total number of events over a large number of time steps should therefore, on average, be underestimated (not accounting for any quantum interference effects)}. This discrepancy is minimized by making the events rare. Furthermore, a high event probability will also lead to an apparent, but unphysical, temporal correlation between such events.

This is typically only an issue in regimes of very high $\chi$ and as the simplest way to ensure sufficient accuracy is decreasing the global time step, it can for the most part be ignored. At the same time this issue can not be completely disregarded, as decreasing the global time step can be highly inefficient. Ultimately, an increased temporal resolution is only necessary for particles experiencing high values of $\chi$, and typically only for a limited period of time as the strong fields giving rise to the high $\chi$ are often well localized in both time and space. The general solution is therefore to implement what is called sub-cycling, wherein the time step is reduced only for the particles for which it is required, as determined by their instantaneous value of $\chi$. This can be performed entirely in the particle pusher, without any changes to the remainder of the PIC loop, and is commonly done by splitting the time step into an integer number of sub-steps for the selected particles.

Although only a fraction of the ensemble is typically contributing to the QED processes, the requirements on the time step will affect the efficiency of the entire event generator, as its accuracy relies on a large rejection-acceptance ratio. This in turn implies that the number of rate computations per accepted event is large. An improvement can be obtained by computing \emph{when} the next emission will occur, instead of \emph{if} it will occur within the current time step. This minimizes the number of expensive computations per event, even allowing it to approach unity without loss of accuracy under certain conditions. Furthermore, this way a complete separation of time scales can be obtained between the QED-event generator and the classical PIC loop, allowing the global time step to be set through consideration of only the usual PIC constraints, such as the stability of the field solver.

\section{\label{sec:method}Method}

\subsection{Memorylessness}
If we assume that events are discrete and independent then the time between events can be described through the exponential distribution. Because the exponential distribution is memoryless, this can be exploited to create an event generator that is independent \changed{of the chosen global time step of the PIC simulation}. This is achieved by computing the time until the next event. If the event is determined to occur within the current time step, it can be performed at the determined time through adaptive sub-cycling. If not, the event can be discarded and the simulation pushed to the next time step. 

We can easily show that this produces consistent sampling of the distribution function, independently of the choice of the time step $\Delta t$, using the cumulative distribution function (CDF) of the exponential distribution, $P_\mathrm{e}(\tstar \leq t) = 1-\exp(-\lambda t)$, where $\lambda$ is the event rate and $\tstar$ is the time until the next event. We further assume that $\lambda$ remains unchanged over one time step. For $t \leq \Delta t$, the probability for an event to occur within a current time step is trivial, as it is entirely determined by the exponential distribution.
\begin{equation}
P(\tstar \leq t; t\leq\Delta t) = P_\mathrm{e}(\tstar \leq t).
\end{equation}
With $t > \Delta t$ however, the CDF instead contains two terms: (1) The probability for an event to occur within a concurrent time step ($P_\mathrm{e}(\tstar \leq \Delta t)$) plus (2) The probability that the event was rejected in a concurrent time step \emph{and} accepted in the succeeding time step. 
\begin{equation}
\begin{aligned}
P(\tstar \leq t; t>\Delta t) &= P_\mathrm{e}(\tstar \leq \Delta t) + P_\mathrm{e}(\tstar > \Delta t)P_\mathrm{e}(\tstar \leq t-\Delta t) \\
&= 1-\e^{-\lambda \Delta t} + \e^{-\lambda \Delta t}(1-\e^{-\lambda (t - \Delta t)}) \\
&= 1-\e^{-\lambda t} = P_\mathrm{e}(\tstar \leq t)
\end{aligned}
\end{equation}
We thus have that $P(\tstar \leq t) = P_\mathrm{e}(\tstar \leq t)$ regardless of the value of $\Delta t$. As such, the choice of time step may, with regard to the event generator, simply be seen as arbitrary synchronization points, allowing us to completely decouple any time step requirement due to particle events from the constraints due to the fields.

\begin{figure}[t!]
    \centering
    \includegraphics[width=0.50\linewidth]{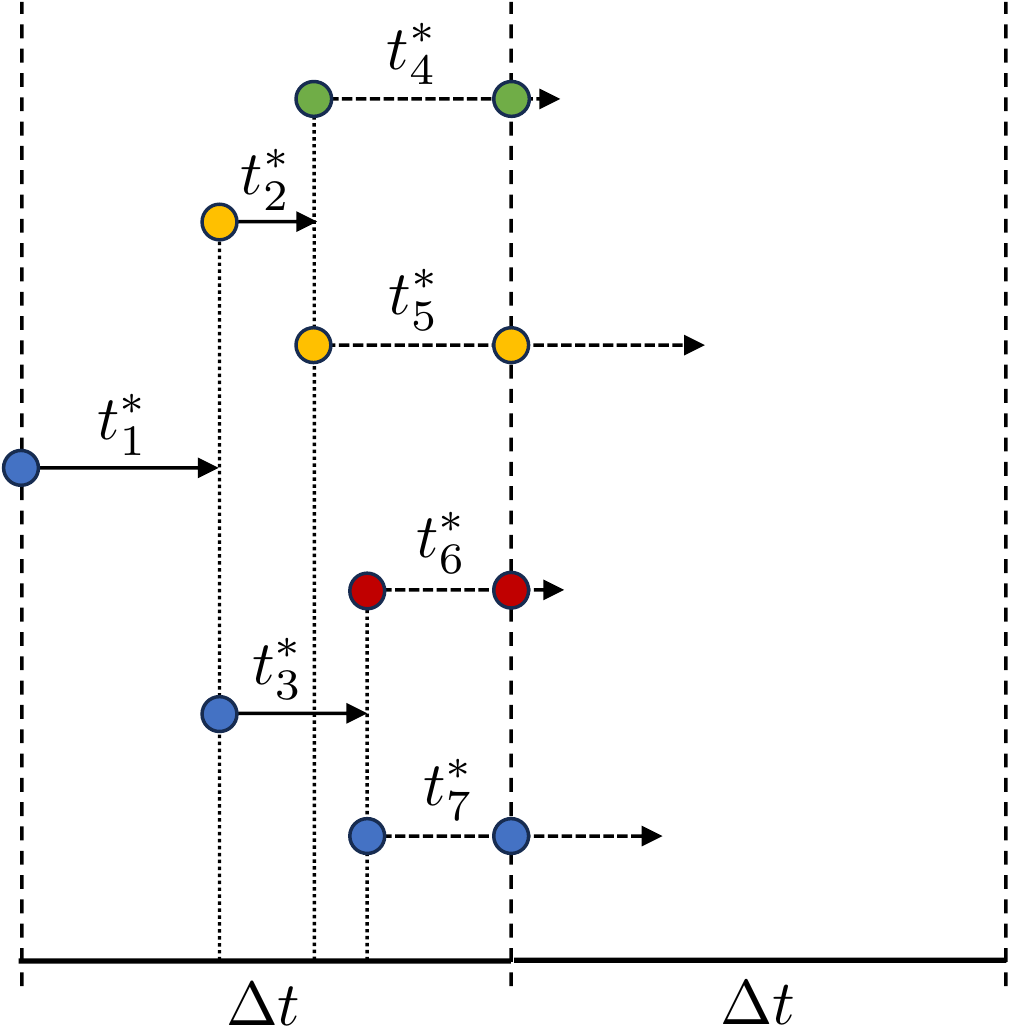}
    \caption{\changed{Graphical representation over how the sub-cycling scheme works. The time until the next event $\tstar$ is computed for each particle separately (colored circles). If the event is computed to occur within the current global time step ($\Delta t$) the particle is pushed to that point in time, after which the event is performed (solid arrows). If the event is to occur outside the current global time step the particle is instead pushed to the end of the global time step, and the event is discarded (dashed arrows). For simplicity, every event is here depicted as creating one more particle.}}
    \label{fig:sketch}
\end{figure}

This important property leads to two significant improvements in comparison with the general approach \cite{ridgers.jcp.2014, duclous.ppcf.2011, tamburini.arxiv.2023}. Firstly, we do not need to impose the smallness on the time step to integrate the optical depth. Instead, at each time step we sample events for each particle, and if the event does not occur on the current step, we simply move the particle for the whole time step and discard the event. Otherwise, we move the particle during the fraction of the time step corresponding to the optical depth, to the position where secondary particles are produced. This procedure ensures temporal locality, i.e. all event processing is performed within a single time step and no state (e.g. an optical depth) needs to be saved between time steps. Secondly, after the event we can sample \changed{for new events from both} initial and secondary particles and move them until all particles move to the end of the time step. \changed{A graphical sketch of how this sub-cycling works is shown in Figure~\ref{fig:sketch}.} Thus, we can account for multiple events during a single time step, which also relax time constraints on a time step in the case of a prolific secondary particles production.

\subsection{Determining the next event}
\label{sec:NextEvent}
For both Compton scattering and Breit-wheeler pair production the event rate is simply determined by the total rate $\lambda = W(\chi) = \int (\rd W/\rd E) \,\rd E$. Knowing that the time between emissions $\tstar$ is governed by the exponential distribution it can, through inverse sampling, be obtained as $\tstar = -\log(r)/W(\chi)$, %-\frac{\log(1-r)}{W(\chi)},
where $r\sim U(0,1)$ is a uniformly distributed random number. This process is equivalent to determining the optical depth of a particle.

\subsection{Photon emission}
The partial rate of nonlinear Compton scattering is given by \cite{nikishov1964quantum, berestetskii1982quantum, nikishov1967pair, baier1967quantum}:
\begin{equation}\label{eq:photon-partial-rate}
\frac{\rd W_\gamma}{\rd\energy} = \frac{\sqrt{3}}{2\pi}\frac{\alpha}{\tau_C}\frac{\chi}{\gamma} \left[ \frac{(1-\energy)}{\energy}F_1(z) + \energy F_2(z) \right], \quad 
z = \frac{2}{3\chi}\frac{\energy}{1-\energy},
\end{equation}
where $\energy = \hbar\omega/\gamma mc^2$ is the photon energy normalized to the energy of the radiating particle, $\hbar$ is the Planck constant, $\omega$ is the photon frequency, $\gamma$ is the Lorentz factor, $ m$ is the electron mass, $c$ is the speed of light, $\alpha$ is the fine structure constant and $\tau_C$ is the Compton time. $F_1(x) = x\int_x^\infty K_{5/3}(x^\prime) \rd x^\prime$ and $F_2(x) = xK_{2/3}(x)$ are the \emph{first and second synchrotron function}, respectively.

\subsubsection{Event rate}
\changed{Consider the total rate of photon emission:
\begin{equation}\label{eq:full-compton-spectrum}
W_\gamma(\chi) = \frac{\sqrt{3}}{2\pi} \frac{\alpha}{\tau_C} \frac{\chi}{\gamma} \int_0^1 \left( \frac{(1-\energy)}{\energy}F_1(z) + \energy F_2(z) \right) \rd\energy 
\end{equation}}
As described above, the main idea of the new method is estimating \changed{the time} $\tstar$ after which the photon is emitted. In order to achieve this, the \changed{Compton rate $W_\gamma(\chi)$} is approximated by a piecewise function over $\chi$, the local trend is identified and refined using a polynomial approximation (intervals and local trends are given in \ref{Appendix:ComptonRate}). The tables specifying intervals over $\chi$ and the local trends are provided in the supplementary materials section for a more compact narration. The employment of polynomials up to the $9^{th}$ order allows us to achieve a relative precision of $10^{-3}$, which will be demonstrated below. \changed{Polynomial values can be computed using the Horner scheme \cite{horner1819xxi}, based on Fused Multiply-Add (FMA) instructions, computing multiplication and addition ($a \cdot b+c$) in one operation with higher accuracy and performance.} As a result, these computations are relatively lightweight for modern CPUs. We note that the proposed approach allows for even higher precision, but that would require additional computations and would be significantly slower. We believe that the proposed parameters of approximation provide an optimal compromise between accuracy and computational efficiency, while accuracy remains at least the same as in the prior scheme \cite{gonoskov.pre.2015}.

\subsubsection{Determining the energy (Compton)} \label{sec:comptonCDF}
The energy can be obtained through inverse transform sampling of the partial rate, requiring the computation of the cumulative distribution function
\begin{equation}
    \cdf_\gamma(\chi, \energy) = \frac{W_\gamma(\chi, \energy)}{W_\gamma(\chi)},
\end{equation}
where $W_\gamma(\chi, \energy) = \int_0^\energy \frac{\rd W}{\rd \energy^\prime}(\chi,\energy^\prime) \,\rd\energy^\prime$.

In order to obtain the required value of $\energy$ the inverse function $\cdf_\gamma^{-1}(\chi, \energy)$ (Compton inverse CDF) must be evaluated for $r\sim U(0, 1)$, for which the method of tabulation of an inverse function was used. \changed{The domain of the inverse function was represented by a mesh as follows. The mesh is uniform over $r_2\in\left[0.05,0.93\right]$ (96 points), inverted logarithmic condensing to $1$ over $r_2\in\left[0.93,1\right]$ (96 points), and logarithmic over $\chi\in\left[1.5^{-30},1.5^{30}\right]$ (61 points). The inverse function was evaluated in the nodes.   
In further computations linear interpolation is used between nodes in the main region and asymptotic approximations are used over $r_2\in\left[0.0,0.05\right]$.} Within this approach an acceptable relative deviation was considered to be below $10^{-2}$ in the specified domain. The typical value was $10^{-3}$. 
\changed{We experimentally found that such a complex distribution of points allows us to approximate the function accurately enough to avoid the problems described in \cite{guo2022improving}.}
Implementation details are given in \ref{Appendix:ComptonCDF}.

\subsubsection{The particle push algorithm with account for QED effects}
\label{sec:particle_push}
The constructed approximations allow us to formulate a new particle push algorithm of charged particles during a single global time step within the main PIC cycle with account for QED effects (algorithm \ref{alg:particle_push}). The algorithm functions as follows. Until the particle exits the loop (lines 3-14), signifying the end of the current time step, the time of emission of the next photon is continuously updated (lines 4-6). If a photon must be emitted within this time step, the particle is pushed to coordinates corresponding to the time of emission (line 11), after which a photon is emitted (line 14) with the according energy (line 13). Otherwise, the particle is pushed to coordinates corresponding to the end of the time step (line 8). The emitted photons are pushed using an algorithm described in Section~\ref{sec:photon_push}. \rev{Note that $\chi$, $rate$, and $\tstar$ are updated at the beginning of each time sub-step, while the fields are assumed to be constant throughout the global time step in accordance with the usual assumptions of the PIC method.}

\begin{algorithm}
\caption{The particle push algorithm with account for QED effects (for charged particles)}\label{alg:particle_push}
\begin{algorithmic}[1]
    \Procedure{ParticlePush}{$particle, E, B, dt$}
    \State $t = 0$
    \While{$t < dt$}
        \State \rev{$Update$} $\chi$
        \State $rate = ComptonRate(\chi)$
        \State $\tstar = -log(random()/rate)$
        \If{$t+\tstar > dt$}
            \State \changed{$Pusher(particle, E, B, dt-t)$}
            \State $t=dt$  
        \Else
            \State \changed{$Pusher(particle, E, B, \tstar)$
            \State $t=t+\tstar$
            \State $invCDF = ComptonInvCDF(\chi, random())$
            \State $EmissionPhoton(particle, invCDF)$}
        \EndIf
    \EndWhile
    \EndProcedure
\end{algorithmic}
\end{algorithm}

\subsection{Photon decay into an electron-positron pair}
\subsubsection{Event rate (Breit-Wheeler)}
The partial rate of Breit-Wheeler pair production is given by \cite{berestetskii1982quantum}:
\begin{equation}\label{eq:pair-partial-rate}
\frac{\rd W}{\rd\energy_e} = \frac{\sqrt{3}}{2\pi} \frac{\alpha}{\tau_C} \frac{mc^2\chi_\gamma}{\hbar\omega}  \left[ (\energy_e - 1)\energy_e F_1(z_p) + F_2(z_p) \right],
\end{equation}
where $z_p = \frac{2}{3\chi_\gamma (1-\energy_e)\energy_e}$. This expression shows the probability of decay of a photon into an electron-positron pair with energies $\energy_e\hbar\omega$ and $(1-\energy_e)\hbar\omega$, respectively.

Similar to the way photon emission is handled, the time of photon decay is determined from the \changed{total pair production rate:
\begin{equation}\label{eq:full-pair_spectrum}
W_p(\chi_\gamma) = \frac{\sqrt{3}}{2\pi} \frac{\alpha}{\tau_C} \frac{mc^2\chi_\gamma}{\hbar\omega} \int_0^1 \left( (\energy_e - 1)\energy_e F_1(z_p) + F_2(z_p) \right) \rd\energy_e.
\end{equation}
By computing the Breit-Wheeler rate $W_p(\chi_\gamma)$ the value of $\tstar$, at which point in time the event occurs, can be determined in accordance with Sec.\ref{sec:NextEvent}.} For fast computation a piecewise function with identified local trends and refinement using polynomials is used (intervals and local trends are given in \ref{Appendix:BWRate}). The demanded precision is the same as in the case of photon emission.

\subsubsection{Determining the energy (Breit-Wheeler)}
\label{sec:BWCDF}
In order to obtain the dimensionless energy of the generated electron $\energy_e$ (the energy of the positron is then $1-\energy_e$) the same approach as in Sec.\ref{sec:comptonCDF} is used.
\begin{equation}
    \cdf_p ( \chi_\gamma, \energy_e) = \frac{W_p(\chi_\gamma, \energy_e)}{W_p(\chi_\gamma)},
\end{equation}
where $W_p(\chi_\gamma, \energy_e) = \int_0^{\energy_e} \frac{\rd W}{\rd \energy_e^\prime}(\chi_\gamma,\energy_e^\prime) \,\rd\energy_e^\prime$.

\changed{In order to evaluate the inverse function $\cdf_p^{-1}(\chi_\gamma, \energy_e)$ (Breit-Wheeler inverse CDF) a mesh logarithmic over $r_2\in\left[0.0,0.05\right]$ (64 points), uniform over $r\in\left[0.05,0.5\right]$ (64 points), and logarithmic over $\chi\in\left[1.5^{-30},1.5^{30}\right]$ (61 points) is used.} Since the function is symmetric with respect to $r_2=0.5$, the values $r\in\left[0.5,1\right]$ are not used and in this case $\cdf_p^{-1}(\chi_\gamma, r)=1-\cdf_p^{-1}(\chi_\gamma, 1-r)$. Implementation details are given in \ref{Appendix:BWCDF}.

\subsubsection{Particle push algorithm with account for QED effects (for photons)}
\label{sec:photon_push}
\changed{
The algorithm of particle push for photons with account for QED effects by the end of a global time step (algorithm \ref{alg:photon_push}) is similar to the algorithm for charged particles. Unlike the algorithm \ref{alg:particle_push}, we do not need a while loop, since the photon is removed during pair production and no further work is required on it.
If a photon must decay within this time step, the photon is pushed to coordinates corresponding to the time of decay (line 10), after which an electron-positron pair is generated (line 13) with the according energies (line 12) and the photon is deleted (line 14). 
Otherwise, the photon is pushed to coordinates corresponding to the end of the time step (line 7). 
The generated particles are pushed using the algorithm described above in Sec.\ref{sec:particle_push}.
}

\begin{algorithm}
\caption{Particle push algorithm with account for QED effects (for photons)}\label{alg:photon_push}
\begin{algorithmic}[1]
    \Procedure{PhotonPush}{$particle, E, B, dt$}
    \State $t = 0$
    \State \rev{$Update$} $\chi_\gamma$, $\upsilon$
    \State $rate = BreitWheelerRate(\chi_\gamma)$
    \State $\tstar = -log(random()/rate)$
    \If{$\tstar > dt$}
        \State $particle.position \mathrel{+}\mathrel{=} dt \cdot \upsilon$
        \State $t=dt$  
    \Else
        \State $particle.position \mathrel{+}\mathrel{=} \tstar \cdot \upsilon$
        \State $t=\tstar$ 
        \State $invCDF = BreitWheelerInvCDF(\chi_\gamma, random())$
        \State $PairProduction(particle, invCDF)$
        \State $Delete$ $particle$
    \EndIf
    \EndProcedure
\end{algorithmic}
\end{algorithm}

\section{\label{sec:validation}Validation}
We here present the accuracy of the implemented functions by comparing the functions to the true values, which are computed to a high accuracy (of the order of $10^{-15}$) through a combination of piecewise integration and special treatment of integrand divergences.

\subsection{Event rates}

\begin{figure}[b!]
    \centering
    \includegraphics[width=0.99\linewidth]{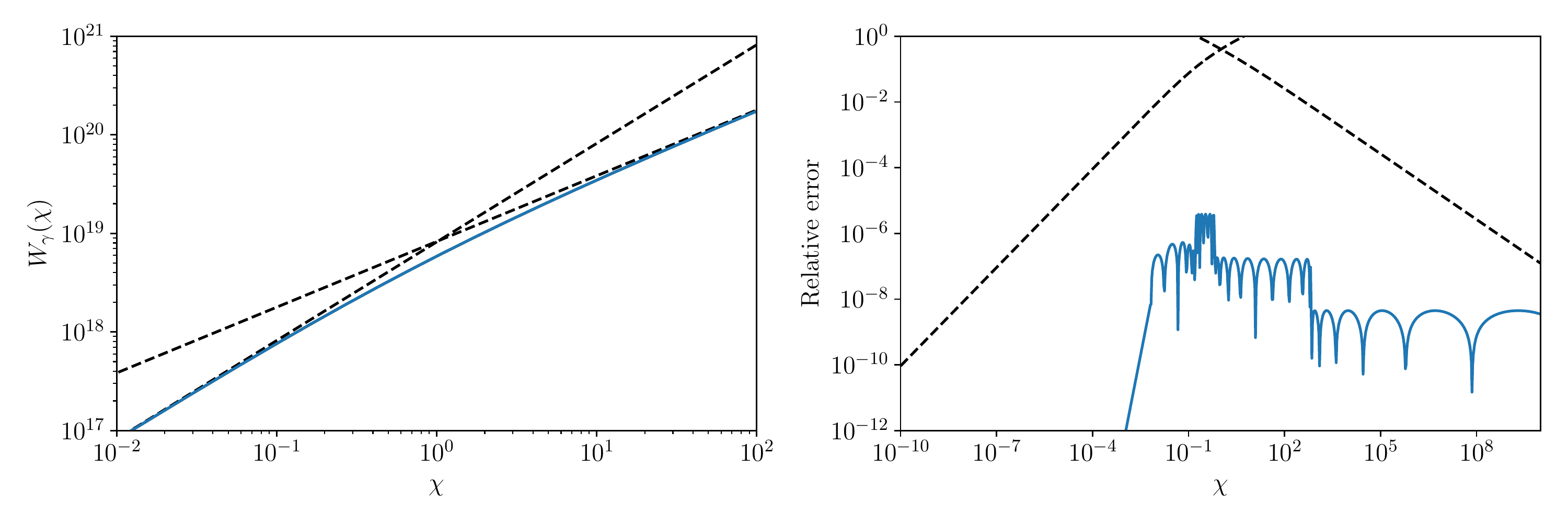}\\
    \includegraphics[width=0.99\linewidth]{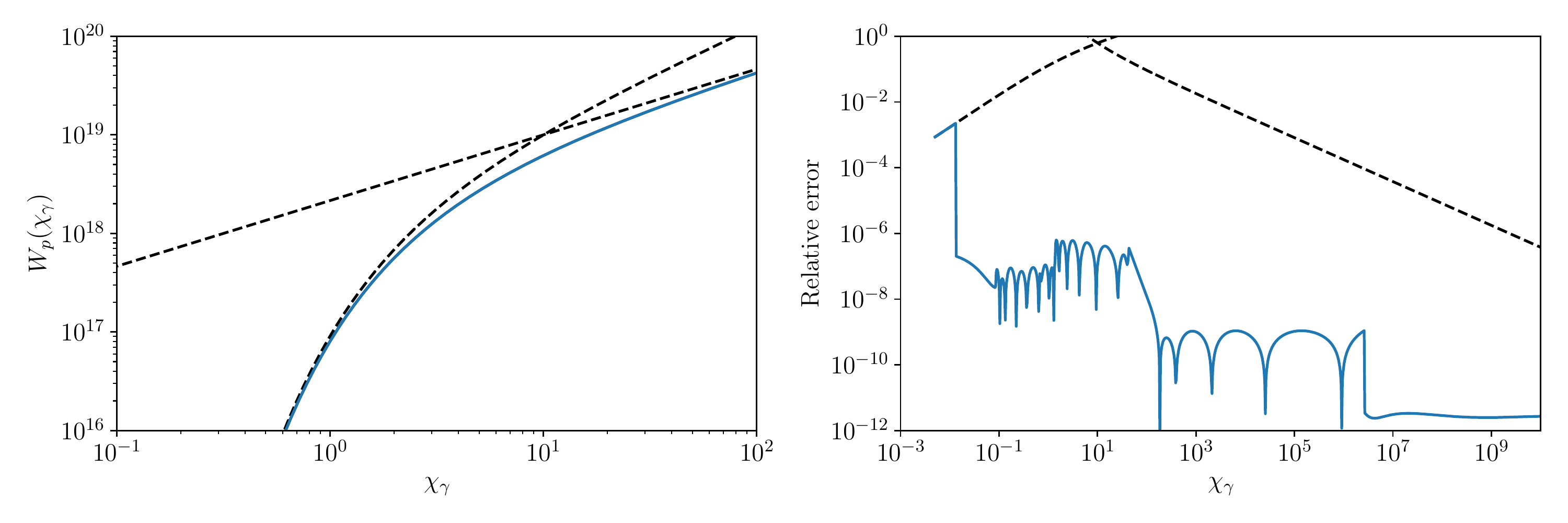}
    \caption{The left panels show the implemented Compton (top, blue) and Breit-Wheeler (bottom, blue) rates, indistinguishable from the true values (solid black), together with the lowest order asymptotic approximations (dashed) as functions of the quantum nonlinearity parameter $\chi$. The right panels show the relative errors of the implemented functions as well as the relative errors of the asymptotic approximations.}
    \label{fig:rates}
\end{figure}

The implemented Compton rate, $W_\gamma(\chi)$, defined in equation~\ref{eq:full-compton-spectrum}, and its relative accuracy are presented in Figure~\ref{fig:rates} and compared with the lowest order asymptotic approximations. The implemented function shows a relative error of less than $2\times10^{-6}$ across all values of $\chi$. Similarly, the implemented Breit-Wheeler rate, $W_p(\chi_\gamma)$, defined in equation~\ref{eq:full-pair_spectrum}, is also presented in Figure~\ref{fig:rates}. The implemented Breit-Wheeler function shows a relative error of less than $10^{-6}$ for all $\chi_\gamma \gtrsim 10^{-2}$. For $\chi_\gamma \lesssim 10^{-2}$ the relative error increases to around $10^{-3}$, but as the Breit-Wheeler rate (as well as the absolute error) in this domain is less than $10^{-70}\,\text{s}^{-1}$, any such errors are computationally irrelevant.

\begin{figure}[b!]
    \centering
    \includegraphics[width=0.49\linewidth]{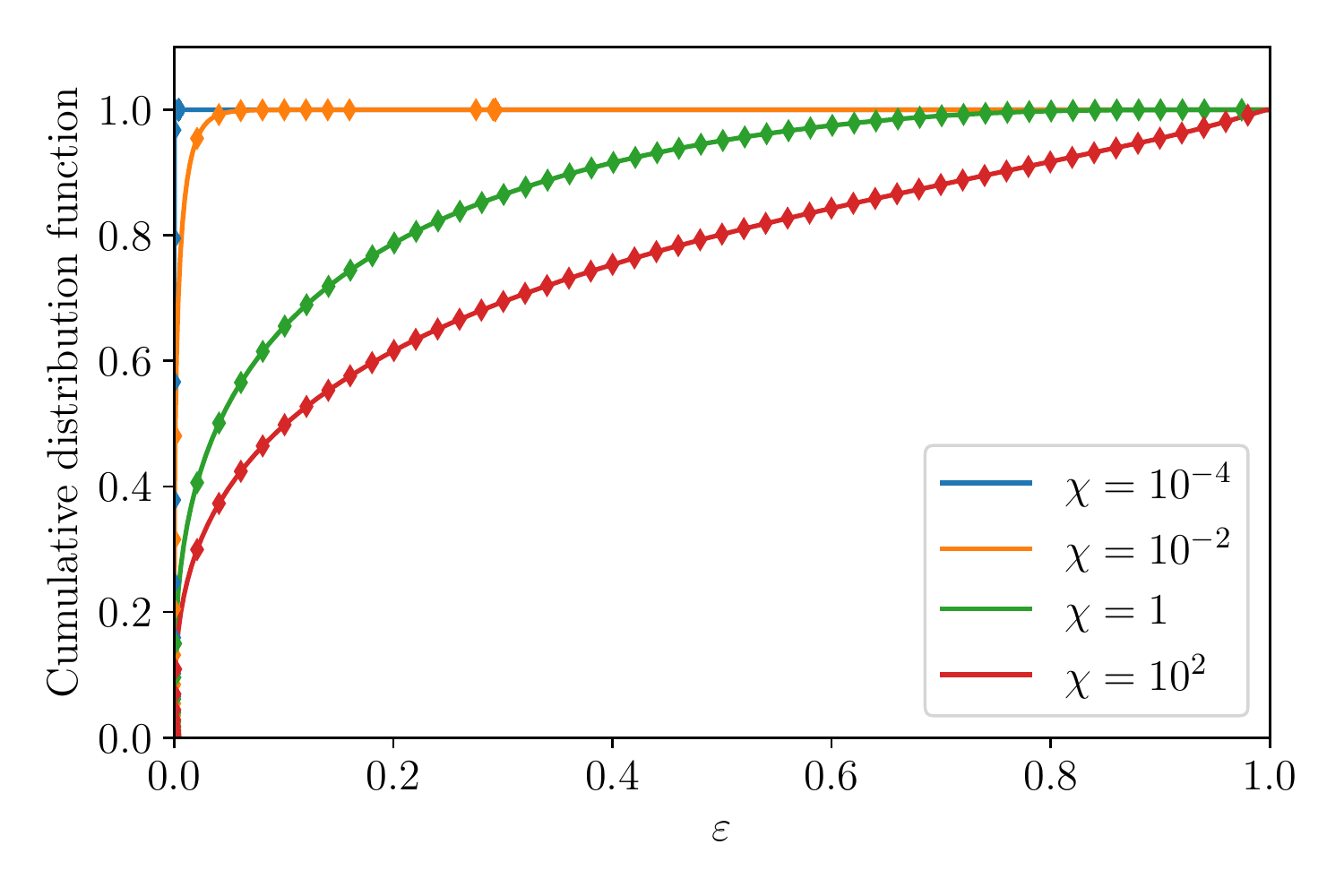}
    \includegraphics[width=0.49\linewidth]{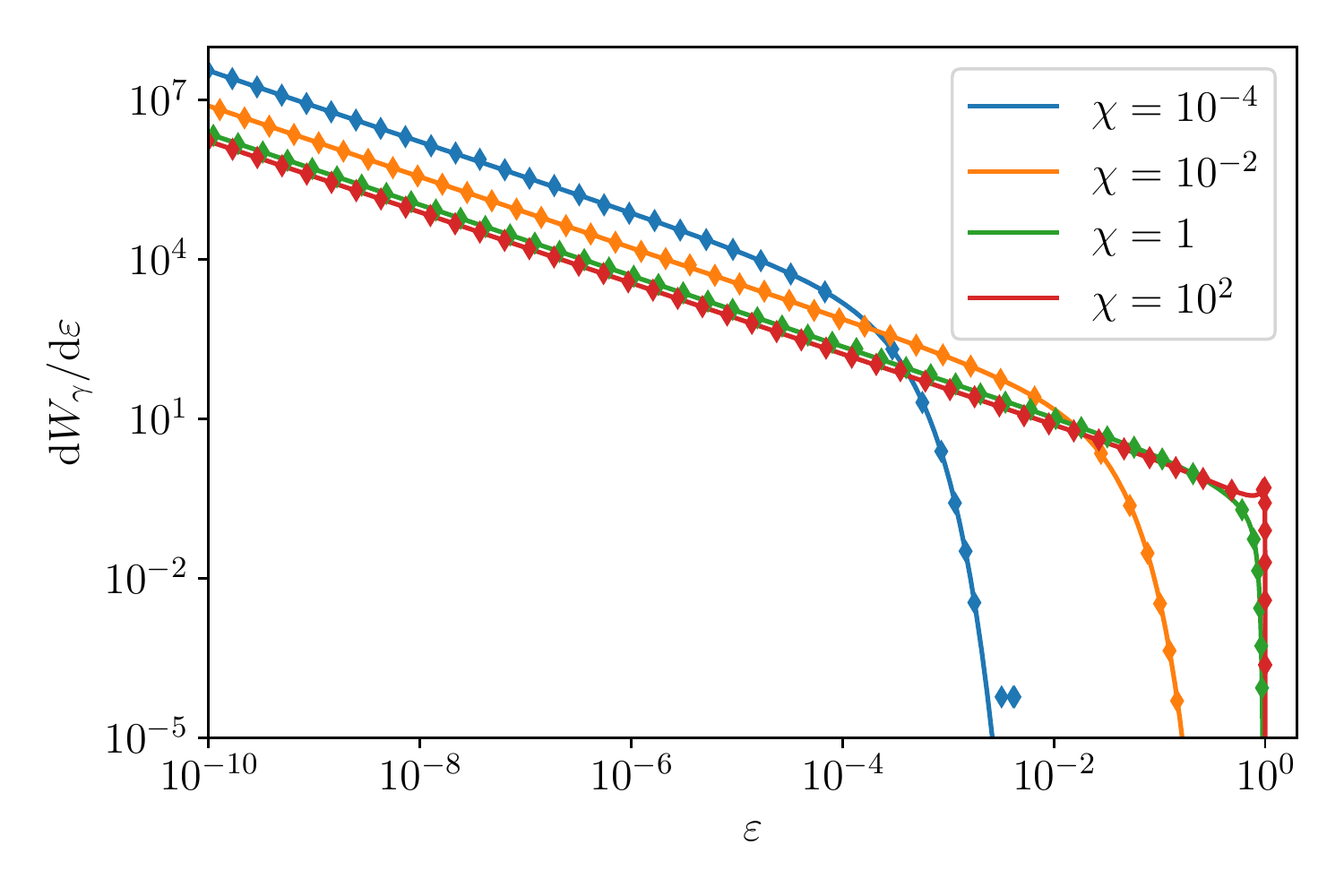}\\
    \includegraphics[width=0.49\linewidth]{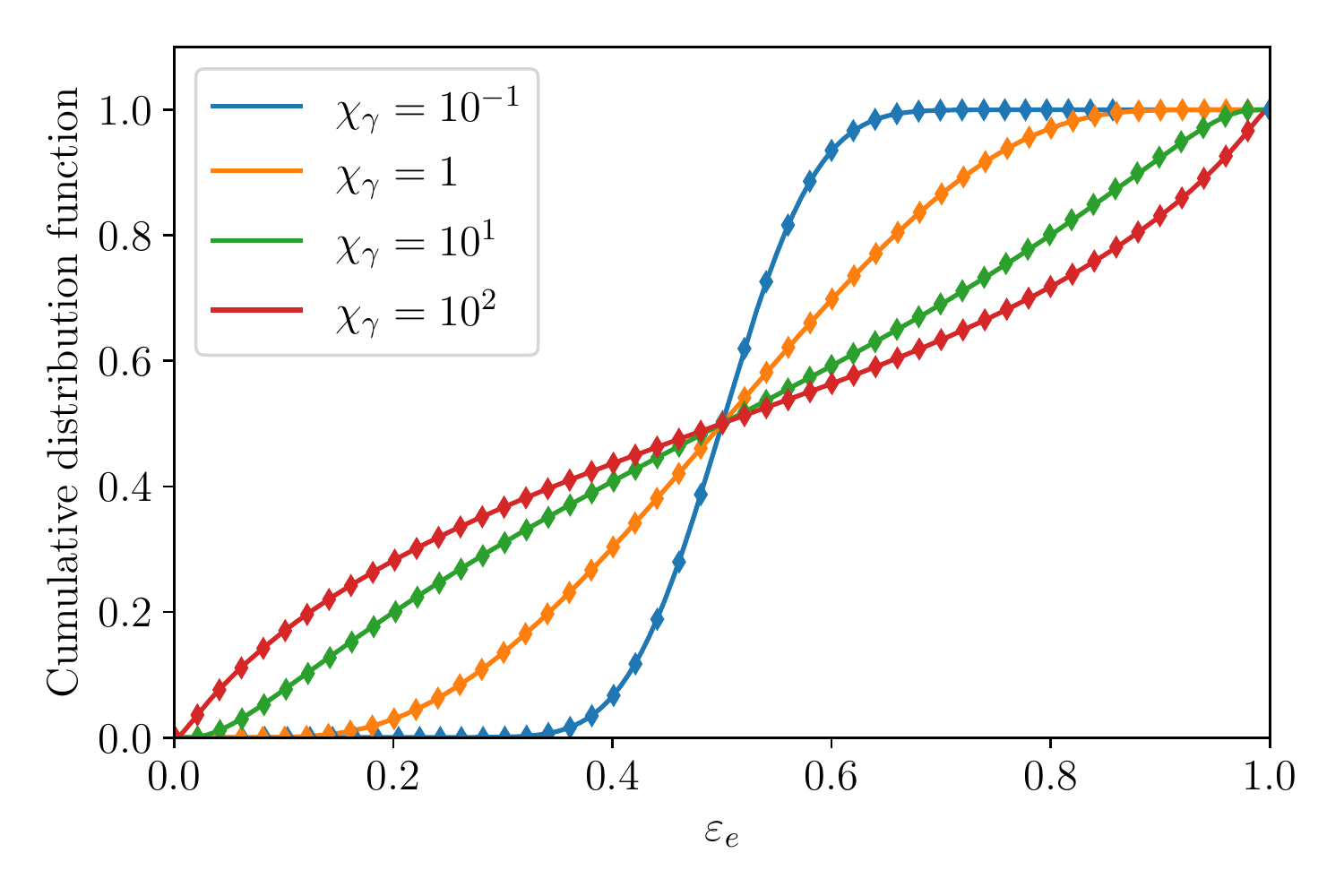}
    \includegraphics[width=0.49\linewidth]{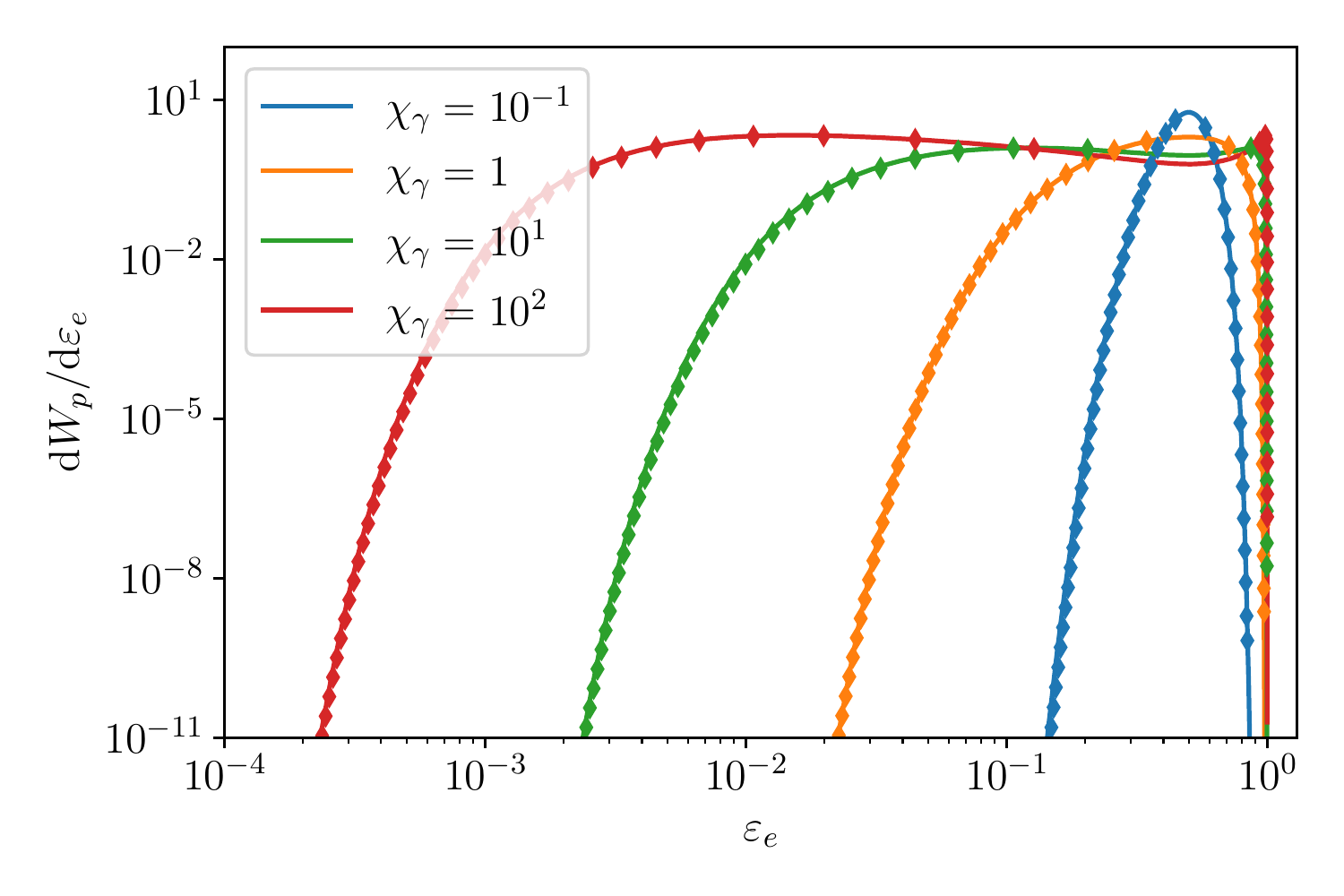}
    \caption{The left panels show the implemented Compton (top, markers) and Breit-Wheeler (bottom, markers) inverse CDFs as functions of energy, compared against the true values (solid lines) for different $\chi$ and $\chi_\gamma$. The right panels show the corresponding number spectra (partial rate) resulting from the inverse CDFs.}
    \label{fig:cdf_spectrum}
\end{figure}

\subsection{Spectra}
The particle spectra are defined through the derivative of the CDF with respect to energy. Because of the involvement of the derivative, the numerical discretization and the two-dimensional nature of the CDF, the accuracy of the produced spectra becomes more difficult to evaluate compared to the accuracy of the rate functions. Care must therefore be taken when interpreting the errors of the inverse CDF, and how it relates to the produced spectra.

With these caveats in mind, the inverse CDF and its resulting number spectra (given by the partial rate) are presented in Figure~\ref{fig:cdf_spectrum} for both the Compton and the Breit-Wheeler processes. Both the CDFs and the spectra show good agreement over a wide range of values of both energy ($\varepsilon$, $\varepsilon_e$) and the quantum nonlinearity parameter ($\chi$, $\chi_\gamma$). Nevertheless, because the CDF becomes non-invertible as $\varepsilon \rightarrow 1$, the relative error contains a divergence in this limit. However, as was the case for the Breit-Wheeler rate in certain regimes, this generally occurs where the absolute errors of the spectra are unquestionably negligible.

\begin{figure}[t!]
    \centering
    \includegraphics[width=0.49\linewidth]{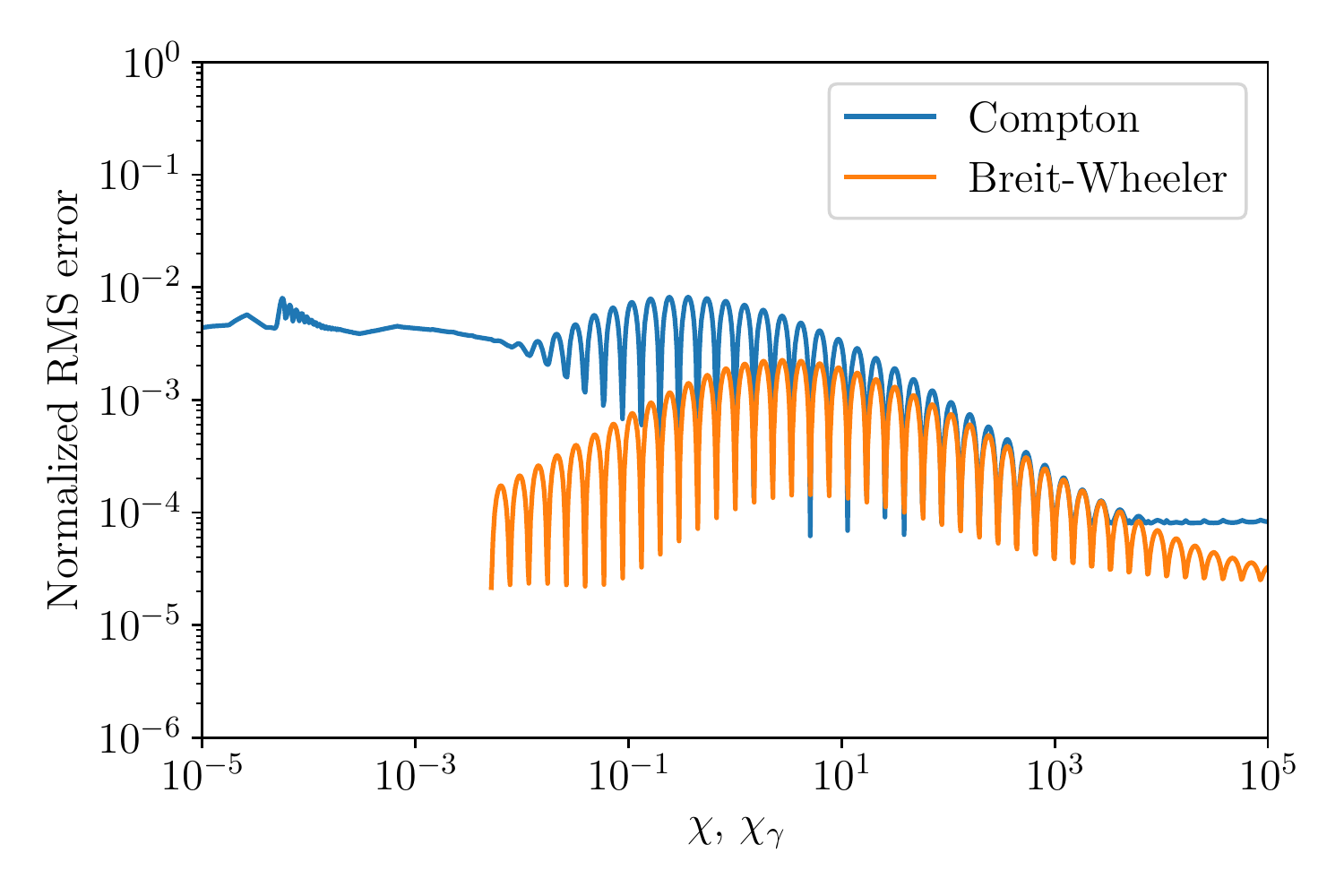}
    \includegraphics[width=0.49\linewidth]{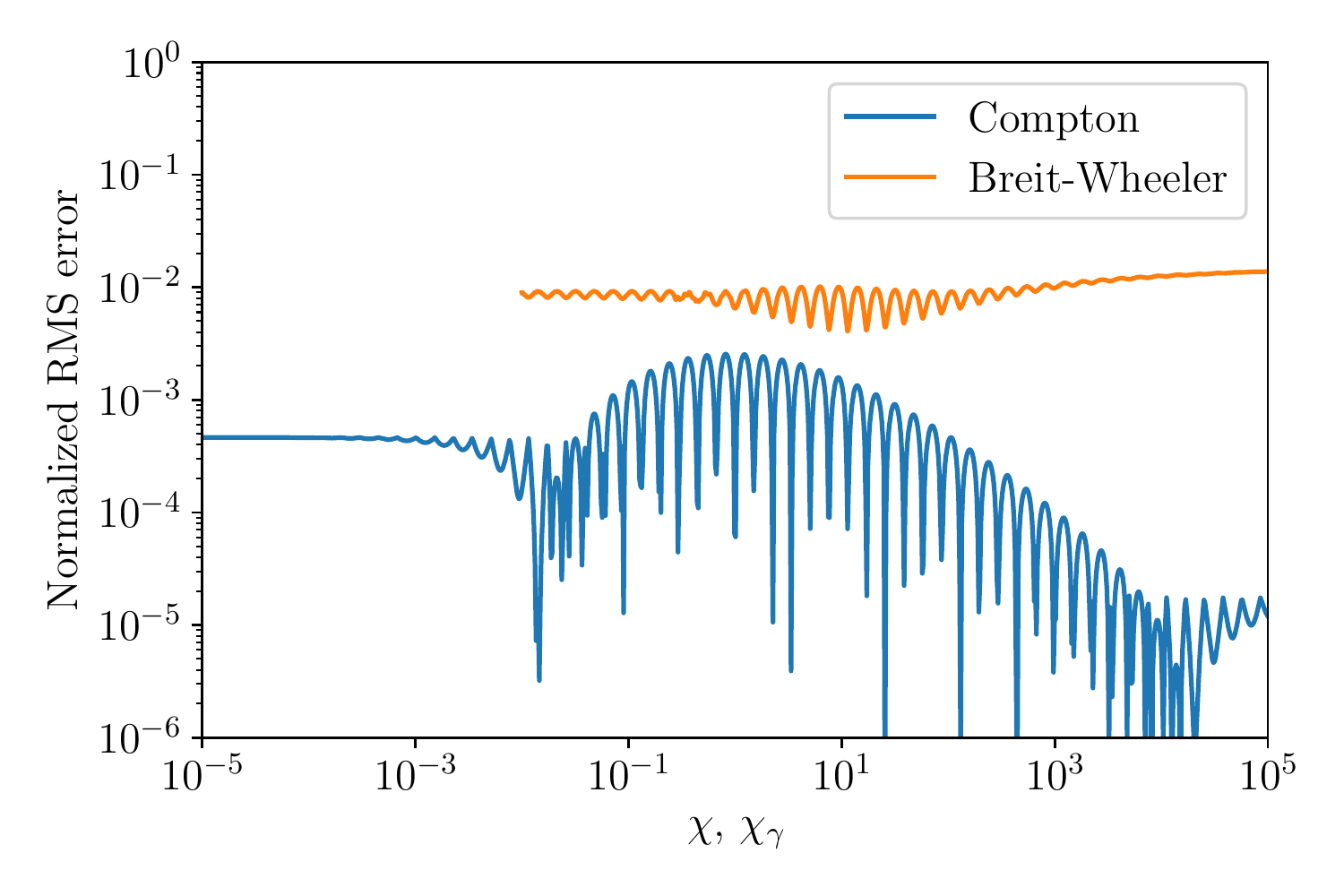}
    \caption{Normalized root mean squared error of the implemented inverse cumulative distribution function (left) and resulting number spectra (right) as functions of the quantum nonlinearity parameter ($\chi$, $\chi_\gamma$) for both the Compton (blue) and Breit-Wheeler (orange) processes.}
    \label{fig:cdf_spectrum_error}
\end{figure}

The normalized root mean squared (RMS) errors of both inverse CDFs and spectra are presented in Figure~\ref{fig:cdf_spectrum_error} as functions of $\chi$ ($\chi_\gamma$). This is computed according to 
\begin{equation}
    \sqrt{\int \big(f(\chi, \varepsilon)-f_0(\chi, \varepsilon)\big)^2\,\rd\varepsilon \left/ \int f_0(\chi, \varepsilon)^2\,\rd\varepsilon\right.},
\end{equation}
where $f$ is the implemented function, and $f_0$ the true function. In general, the inverse CDF and its resulting spectra show typical errors on the order of $10^{-2}$-$10^{-3}$ depending on the regime and process.

\section{\label{sec:verification}Numerical verification}
Here we conclude by testing our implementation on a few benchmark cases and comparing against previously published results.

\subsection{Cascade development}
\begin{figure}[t!]
    \centering
    \includegraphics[width=0.85\linewidth]{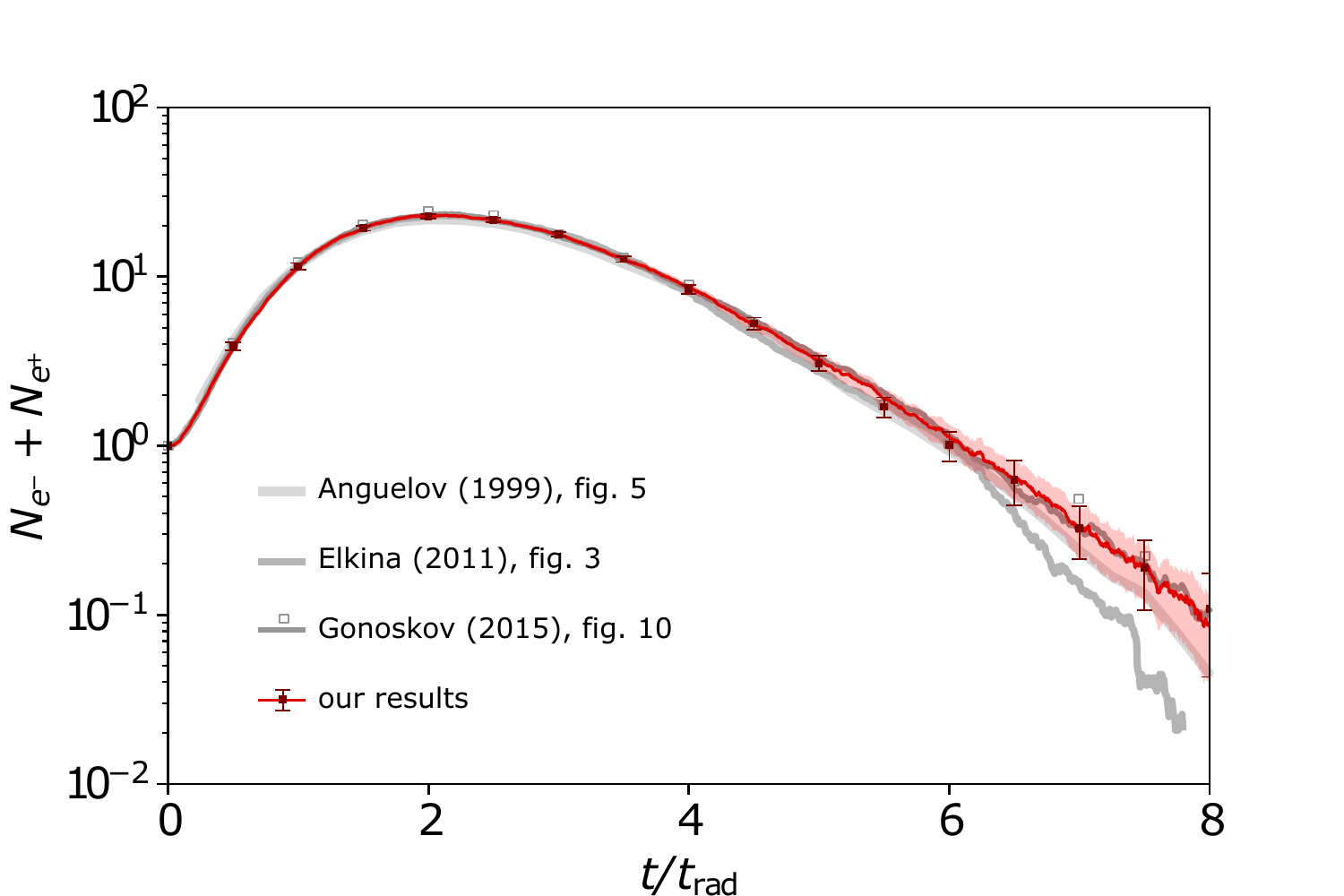}
    \caption{Number of electrons and positrons with energy exceeding $10^{-3}\varepsilon_0$ in a cascade seeded by a single electron with gamma-factor $\gamma_0 = \varepsilon_0/m_ec^2 = 2\times10^{5}$ in a constant magnetic field, transverse to the electron's velocity and of strength $0.2E_S$, presented as a function of time. The results of \cite{anguelov.1999} (light gray), \cite{elkina.prstab.2011} (medium gray) are shown with solid lines. Our results (red) and the results of \cite{gonoskov.pre.2015} (dark gray) are presented for two different choices of time step, $0.5t_\mathrm{rad}$ (markers) and $10^{-3}t_\mathrm{rad}$ (lines). Error bars and the colored area indicate the one sigma standard error of the mean across $1000$ simulations.}
    \label{fig:cascade}
\end{figure}

In this benchmark we simulate the development of a cascade seeded by a single electron with an initial energy of $\varepsilon_0 = \SI{100}{\GeV}$, moving in a strong constant magnetic field ($H_0 = 0.2E_\mathrm{S}$) oriented perpendicularly to the electron's direction of motion. In Figure \ref{fig:cascade} we show the number of electrons and positrons with energy greater than $10^{-3}\varepsilon_0$ as a function of time. The results are averaged over $1000$ simulations and are performed for two different time steps, $10^{-3} t_\mathrm{rad}$ and $0.5 t_\mathrm{rad}$, where $t_\mathrm{rad} = 3.85 \times \gamma_0^{1/3}(E_S/H_0)^{2/3}\hbar^2/(mce^2)$. The benchmark shows excellent agreement between the two choices of time step, as well as with previously published results \cite{anguelov.1999, elkina.prstab.2011, gonoskov.pre.2015}. Some deviations can be seen for $t/t_\mathrm{rad}>6$, however all discrepancies are shown to lie within 1-2 standard deviations.

\subsection{Cascade spectra}
\begin{figure}[ht!]
    \centering
    \includegraphics[width=0.85\linewidth]{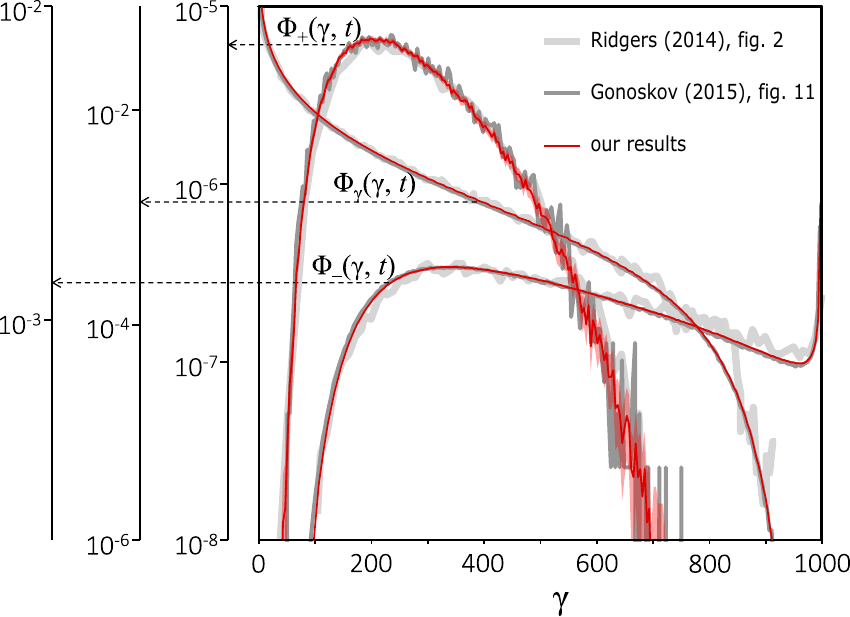}
    \caption{Distribution of electrons $\Phi_-(\gamma, t)$, positrons $\Phi_+(\gamma, t)$ and photons $\Phi_\gamma(\gamma, t)$ as functions of energy, normalized to $mc^2$, after $t = 1\,\text{fs}$ of cascade development. The results of \cite{ridgers.jcp.2014} (light gray), \cite{gonoskov.pre.2015} (medium gray) and our results (red) are shown with solid lines. The colored area indicate the one sigma standard error of the mean across $10^8$ simulations.}
    \label{fig:cascade_spectrum}
\end{figure}

Next, we again simulate a cascade seeded by a single electron in a strong constant magnetic field oriented perpendicularly to the electron's direction of motion, but here with a field strength of $H_0 = 10^{-3}E_\mathrm{S}$ and initial electron energy $\gamma_0 = 1000$. The distribution of electrons $\Phi_-(\gamma, t)$, positrons $\Phi_+(\gamma, t)$ and photons $\Phi_\gamma(\gamma, t)$ are presented in Figure~\ref{fig:cascade_spectrum} as functions of energy and after $1\,\text{fs}$ of cascade development. Our results are averaged over $10^8$ simulation runs and compared with the results of \cite{ridgers.jcp.2014, gonoskov.pre.2015}, showing excellent agreement. 

\subsection{QED cascade in circularly polarized standing wave}
In this section we demonstrate capabilities of the developed method of accounting for QED processes. We consider a model example, which is nevertheless very instructive and widely used for physical insight: a QED cascade in the fields of a standing circularly polarized plane wave \cite{bell.prl.2008,fedotov.prl.2010,bashmakov.pop.2014,kostyukov.pop.2016,grismayer.pre.2017,bashinov.pra.2017,slade.njp.2019}. This wave is homogeneous in directions along both electric and magnetic fields, these fields rotate in time and their amplitudes are locally constant. QED processes are more frequent in the vicinity of the electric field antinode where electrons and positrons quickly gain energy after photon emission and photons have a high probability to decay. In this field configuration, when the QED cascade becomes steady-state, the number of electrons, positrons and photons grows exponentially in time, moreover, radiation losses suppress particle escape from the antinode region \cite{bashinov.pra.2017}. Thus, the considered configuration of fields can lead to likely the most frequent (in comparison with other field structures) calls of functions handling QED processes, and the calculations of these functions may take a significant part of computational time.

We performed series of simulations of a QED cascade in the standing circularly-polarized plane wave with different time steps and wave amplitudes. We analyzed the results and compared the computational performance using different event generators: AEG (the adaptive event generator) \cite{gonoskov.pre.2015}, mAEG (the modified adaptive event generator) \cite{volokitin.jpcs.2020} and FQED (the fast QED-event generator considered in this paper). For simulations with these event generators we used the same configurations. The simulation box was $64\times2\times2$ cells and $2\lambda\times2\lambda\times2\lambda$ along $x$, $y$ and $z$, where $\lambda=0.8$~$\mu$m is the laser wavelength. The time step was varied in range from $T/6000$ to $T/50$, where $T\approx2.67$~fs is the laser period. QED cascade development was considered during $10T$. That period of time is sufficient in order to determine the cascade growth rate. 

\begin{figure}[ht!]
	\centering
	\includegraphics[width=0.99\linewidth]{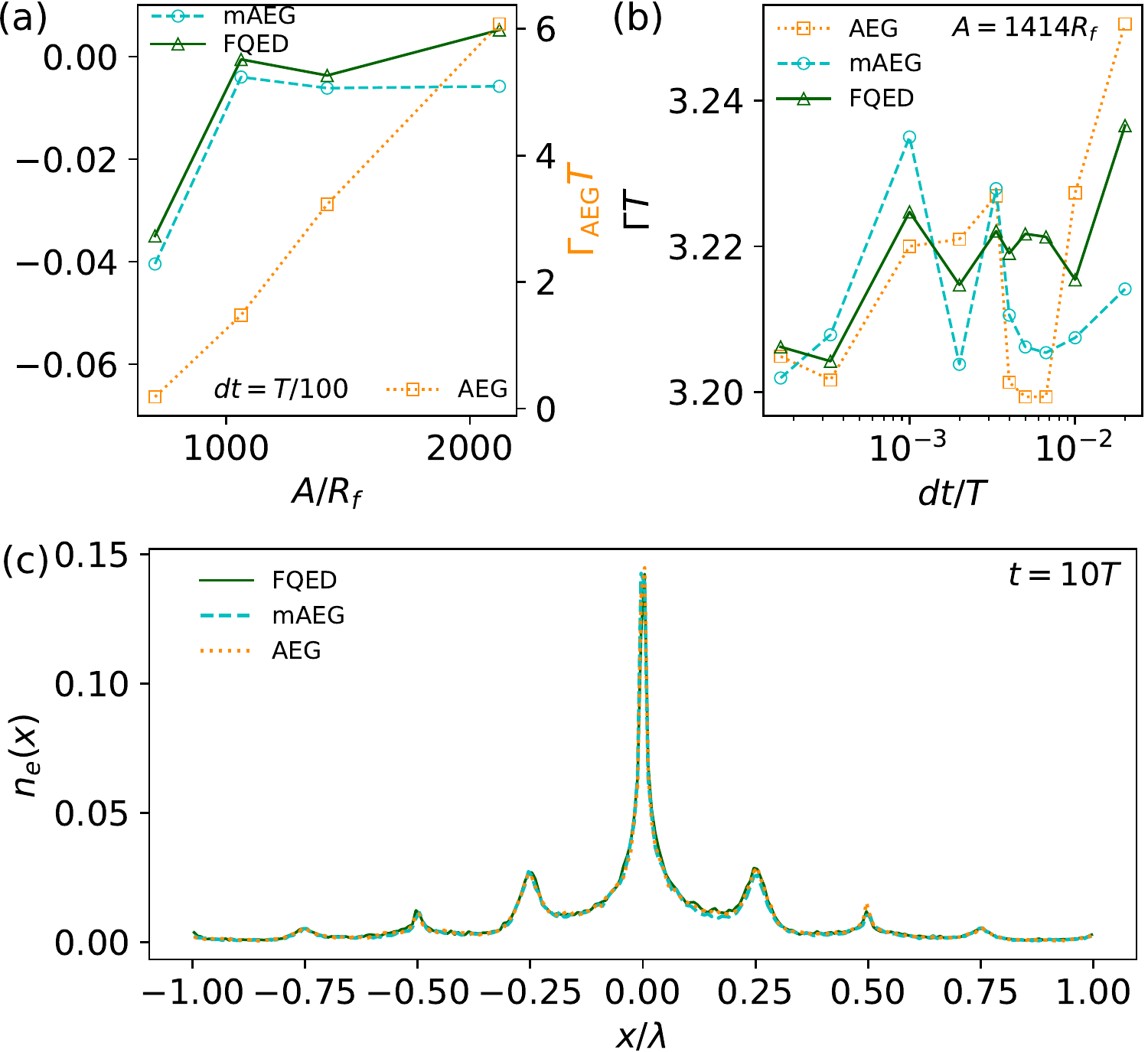}
	\caption{(a) Dependencies of the QED cascade growth rate on field amplitude using different event generators with time step $dt=T/100$. Solid and dashed lines correspond to the relative discrepancy of growth rates $\eta_\Gamma$ (left $y$ axis) calculated with FQED and mAEG with respect to the growth rate $\Gamma_\mathrm{AEG}$ calculated with AEG , represented by the dotted line (right $y$ axis). (b) Growth rates determined with the help of different event generators as functions of the time step at wave amplitude $A=1414R_f$. Markers correspond to performed simulations. (c) Distribution of electron density along the $x$ axis at the last moment of time $t=10T$ in simulations with different QED-event generators.}
	\label{fig:GrRt_CW}
\end{figure}

Fields were set analytically:
\begin{equation}
	\begin{array}{rcl}
		E_y &=& A/\sqrt{2}\times\cos(kx)\sin(\omega t),\\
		E_z &=& A/\sqrt{2}\times\cos(kx)\cos(\omega t),\\
		B_y &=& -A/\sqrt{2}\times\sin(kx)\sin(\omega t),\\
		B_z &=& -A/\sqrt{2}\times\sin(kx)\cos(\omega t),\\
	\end{array}
	\label{eq:CPSW}
\end{equation}
other field components were zero. Amplitude $A$ was considered in the range from $700R_f$ to $2100R_f$, where $R_f = 2\pi mc/eT = 1.3\times10^8$~G or statV/cm. For each amplitude and each time step we performed six simulations: two simulations for each QED-event generator with different seeds for the random number generator (RNG). Results of simulations, as well as their performance at certain physical parameters, were averaged over these two simulations.

Initially electrons and positrons were uniformly distributed in the whole simulation box and the quantity of each was $5\times10^4$. One macroparticle (aggregation of a number of physical particles in PIC-codes) was equal to one real particle. The boundary conditions for particles were periodic due to the plane structure of the continuous wave. In order to avoid memory shortage as a result of QED cascade development we used a global leveling resampling technique \cite{muraviev.cpc.2021}, which does not allow the number of macroparticles of any individual type to substantially exceed a manually specified threshold. The chosen resampling method produces minimal numerical artifacts in the particle ensemble\cite{muraviev.cpc.2021}. The resampling threshold was set such that resampling was run when the number of macroparticles of any given type exceeded $10^6$. 

\begin{figure}[ht!]
	\centering
	\includegraphics[width=0.99\linewidth]{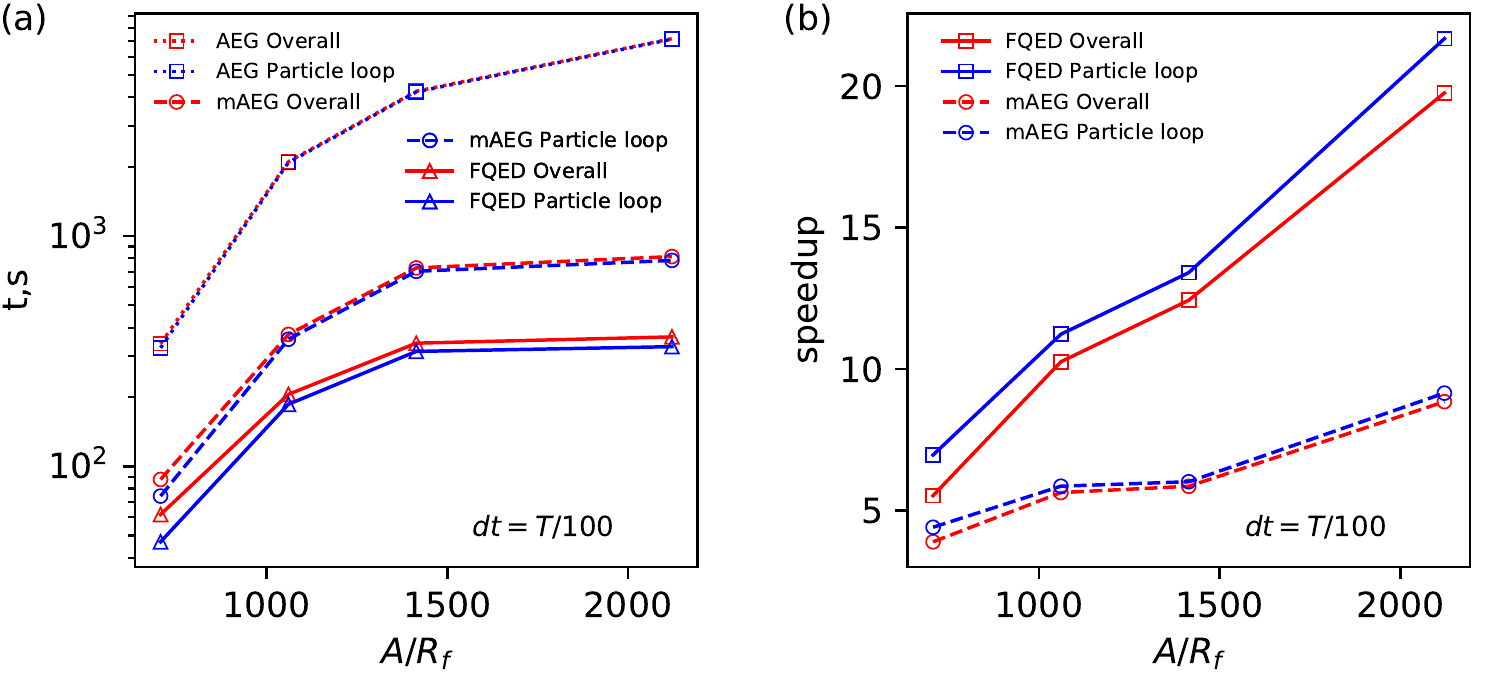}
	\includegraphics[width=0.99\linewidth]{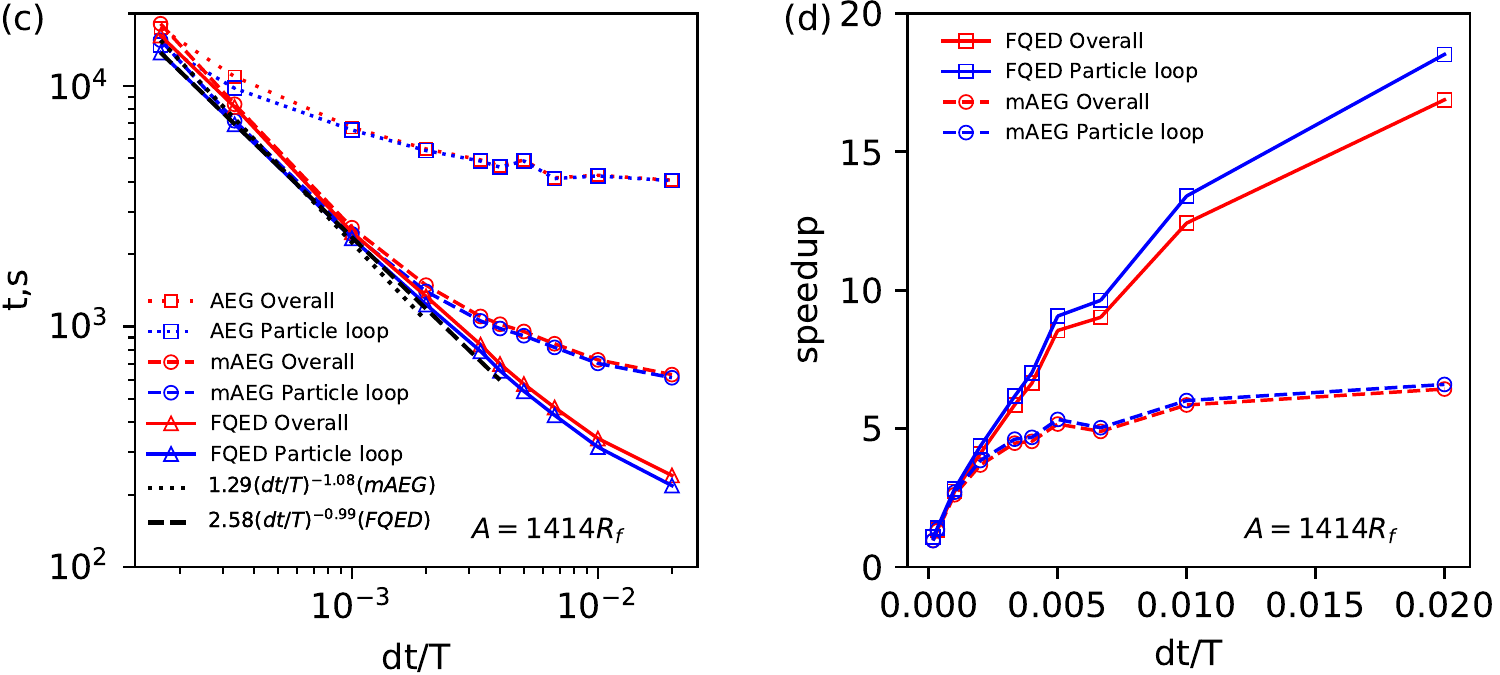}
	\caption{Duration of the particle loop stage and overall duration of the simulation using different event generators (a) as functions of amplitude of the circularly polarized standing plane wave with time step $dt=T/100$ and (c) as a function of time step for wave amplitude $A=1414R_f$. Speedup of the particle loop stage and overall speedup of simulations with mAEG and FQED relative to simulations with AEG as functions of (b) amplitude and (d) time step. Asymptotic behavior of the duration of the particle loop stage with the use of mAEG and FQED event generators are shown by dotted black and dashed black lines, respectively.}
	\label{fig:times_CW}
\end{figure}

For tests of the developed event generator we used 1 two-CPU node of the Joint Supercomputer Center of RAS with Intel Xeon Gold 6248R. We run tests using 1 MPI process and 96 threads.

\textbf{First}, we analyze physical consistency of simulations performed with different event generators. We consider the QED cascade growth rate $\Gamma$ as the main result to compare. In order to determine $\Gamma$ we calculate the total quantity of electrons in the simulation box as a function of time $N_e(t)$ and then $\Gamma=\ln{\left[N_e(10T)/N_e(8T)\right]}/\left(2T\right)$. As reference values of the growth rate we consider values $\Gamma_\mathrm{AEG}$ obtained with the AEG event generator previously used in many studies (for example, see \cite{muraviev.cpc.2021,gonoskov.prx.2017,efimenko.scirep.2018,efimenko.pre.2019,bashinov.pra.2017}).

Figure~\ref{fig:GrRt_CW}~(a) shows that $\eta_\Gamma=\left(\Gamma_\mathrm{mAEG,FQED}-\Gamma_\mathrm{AEG}\right)/\Gamma_\mathrm{AEG}$ (relative discrepancy of growth rates) is about 1\%. Within the amplitude range $A/R_f>1000$, where $\Gamma T>1$,  the relative discrepancy is minimal ($\lesssim1\%$). However, within the lower amplitude range $A/R_f<1000$ for both the mAEG and the FQED event generators $\eta_\Gamma$ is several percent. This increase of $\left|\eta_\Gamma\right|$ can be explained by low values of $\Gamma T<1$ and consequently a larger dispersion of results which may demand averaging over more than two performed simulations with different RNG seeds.

To determine the dependence of $\Gamma$ on time step we choose $A=1414R_f$ at which $\Gamma T>1$ and the dispersion of results is assumed insignificant. For all three event generators fluctuations of the growth rate were within 1\% (average growth rate $\approx3.215T$), see Fig.~\ref{fig:GrRt_CW}~(b). The use of the FQED event generator results in minimal fluctuations within 0.6\%, while AEG leads to the maximal fluctuations around 1\%. The maximal considered time step is $dt=T/50$, since, according to our simulations, for longer time steps the time resolution of particle motion becomes insufficient for the given field amplitude $A=1414R_f$.

Besides growth rate, which is calculated based on the amount of particles in the simulation box, we also show [see Fig.~\ref{fig:GrRt_CW}~(c)] that spatial electron distributions obtained in simulations with different event generators are very close to each other. These distributions are local characteristics and they also confirm that the developed QED-event generators agree nearly perfectly.

\textbf{Second}, we compare the achievable performance of different event generators. The main part of the computations for the considered setup is related to handling of particles as the total duration of each simulation is only slightly longer than the total time spent in the particle loop stage (Fig.~\ref{fig:times_CW}~(a), (c)). Moreover, a substantial part of computations within the particle loop are devoted to QED processes, especially for $A>1414R_f$. An increase of amplitude makes QED processes more probable leading to production of more pairs and photons, substantially increasing both the time spent in the particle loop as well as the overall computational time. However, owing to resampling, which at $A\gtrsim1000R_f$ occurs not only for photons but also for pairs, $t(A)$ is a concave function (see Fig.~\ref{fig:times_CW}~(a)).

Within the considered amplitude range the FQED event generator is the fastest. Its speedup in relation to AEG increases with amplitude and can exceed the factor of 20 (Fig.~\ref{fig:times_CW}~(b)). Another event generator, mAEG, demonstrates qualitatively similar behavior, but speedup is around 2-2.5.

For a given wave amplitude $A=1414R_f$ we also investigate performance at different time steps [Fig.~\ref{fig:times_CW}~(b), (d)]. At time steps $dt\ll T/3000$ all three event generators show very close performance and the time of computations scales as $\sim1/dt$. AEG and mAEG do not need sub-cycling and mainly lines 7-9 of the algorithm \ref{alg:particle_push}, and lines 6-8 of the algorithm \ref{alg:photon_push} of FQED are enabled.

With increasing time step the AEG event generator is the first to have $t$ deviate from the $1/dt$ dependence. At $dt>dt_{AEG}\approx T/3000$ it enables sub-cycling more frequently and as a result at $dt>T/150$ the performance of simulations with this event generator becomes nearly independent on time step.
The QED-event generator mAEG determines sub-steps more precisely \cite{volokitin.jpcs.2020}. So, for this event generator the computational time $t$ scales as $1.2\left(dt/T\right)^{-1.08}$ up to $dt_{mAEG}\approx T/1000$ and at $dt\gtrsim T/100$ the slope of the function $t(dt)$ substantially decreases.

The event generator FQED shows the best performance among the considered event generators. The threshold time step at which $t(dt)$ changes its dependence is $dt_{FQED}\approx T/500$, and deviation from this dependence with increasing $dt$ is quite slow. The maximal speedup is achieved at the maximal possible dt: performance of simulations with FQED can be more than an order of magnitude better than that with AEG and up to an order of magnitude better than that with mAEG.

Owing to properties of particle motion \cite{bashinov.pra.2017}, the characteristic time of QED processes of the considered problem is roughly $t_\mathrm{QED}\sim1.5T\left(R_f/A\right)^{0.645}\approx T/70$. Having compared $dt_{AEG, mAEG, FQED}$ with $t_{QED}$ we conclude that AEG uses excessively short time steps during sub-cycling ($t_{QED}/dt_{AEG}>40$). The more precise estimate of sub-steps for sub-cycling implemented in mAEG ($t_{QED}/dt_{mAEG}\approx14$) significantly improves performance, but the FQED event generator based on both the optical depth methodology and sub-cycling yields the best performance. The necessity of sub-cycling for this event generator (lines 11-14 of the algorithm \ref{alg:particle_push}, and lines 10-14 of the algorithm \ref{alg:photon_push}) is evident at time steps closest to $t_{QED}$ ($t_{QED}/dt_{FQED}\approx7$), and the optical depth methodology allows creating particles or photons with a minimal number of function calls required in order to determine QED rates and generate random numbers. 

To sum up, the developed FQED event generator of QED processes presents advantages compared to both mAEG and AEG: it has approximately the same accuracy but speeds up performance by an order of magnitude or more. The greatest benefit of FQED is observed when the QED rates are greatest (field amplitudes $\gtrsim 1000R_f$) and the time step is quite long. However, it is important to note that there may be limiting factors (computational or physical) that prohibit the use of long time steps at the considered field amplitudes and thus hinder computational performance. For example, particle motion or plasma oscillations may demand (within the employed numerical schemes) very detailed time resolution, with which the speedup provided by FQED is not so impressive, even for larger field amplitudes.

\subsection{QED cascade in two colliding tightly focused laser beams}

Although the developed event generator FQED demonstrated exceptional performance in the previous example, it is important to note that the benefit of using this method is problem-dependent. The proportion of the particle ensemble that takes part in QED processes is crucially important, as well as how strong the electromagnetic fields are, and what time resolution is sufficient. These aspects can be interconnected. For example, if we aim to enhance QED processes, laser intensity could be increased by using tighter laser focusing. This narrows the strongest-field region, making it more inhomogeneous. As a result, particles may escape the strong field region more quickly making a smaller portion of the particle ensemble continuously engaged in QED processes. Additionally, stronger field amplitudes may require shorter time steps. As a result, the interplay of different factors can diminish some of the benefits of FQED.

\begin{figure}[ht!]
	\centering
	\includegraphics[width=0.99\linewidth]{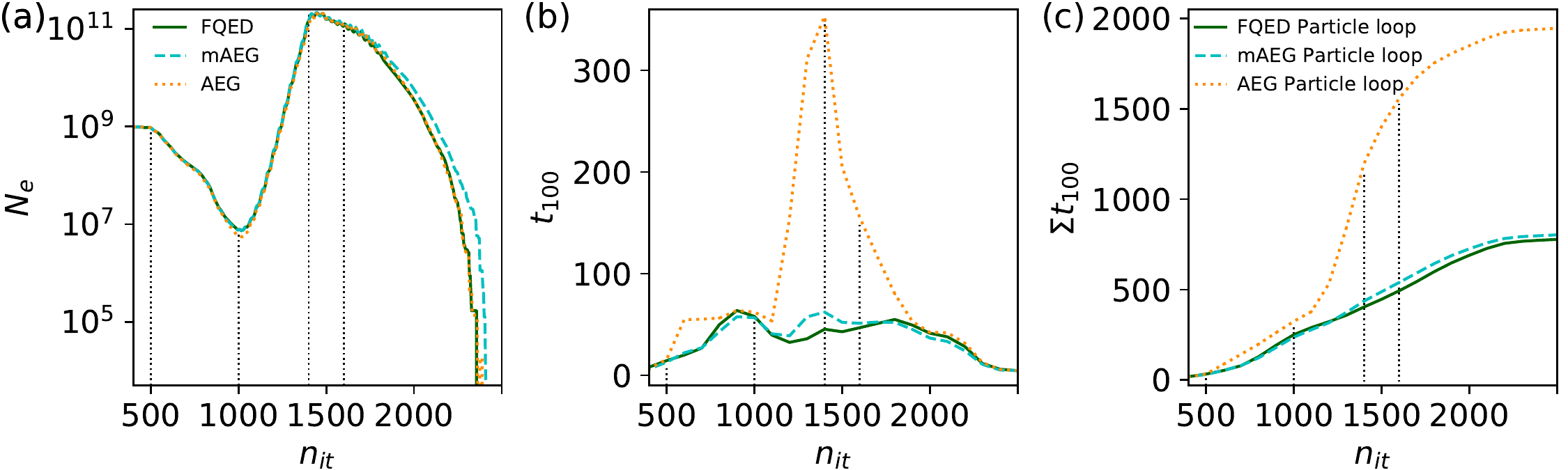}
	\caption{Results of irradiation of a spherical target of $1\lambda$ radius with initial electron density $5n_{cr}$ by two counter-propagating linearly polarized tightly focused laser pulses with power 25~PW and duration of 30~fs. (a) Quantity of electrons $N_e$ in the vicinity of the plane $y=0$ ($-0.03\lambda<y<0.03\lambda$) as function of iteration number $n_{it}$. (b) Time duration demanded for computation of 100 iterations $t_{100}$ depending on iteration number. (c) Cumulative sum of $t_{100}$ to a certain iteration $n_{it}$ ($\Sigma t_{100}$). Legends for (b) and (c) are identical. Vertical black dotted lines separates different stages of laser-plasma interaction.}
	\label{fig:Real1}
\end{figure}

In this section we consider one of the basic schemes of expected experiments devoted to QED cascades using counter-propagating multipetawatt laser pulses (see, for example, \cite{nerush.prl.2011,mironov.qe.2016,artemenko.pre.2017,tamburini.sr.2017,jirka.sr.2017}). Thanks to remarkable progress in laser science and technology such laser pulses are now almost in operation \cite{radier.hplse.2022}. In contrast to the example in the previous section here we consider a more realistic numerical setup. Two identical counter-propagating multipetawatt laser beams with focusing angles equal to $45^\circ$ irradiate a seed hydrogen-like target in the form of a sphere with radius $1\lambda$ located in the mutual focus of these beams. Electron density of the target for numerical tests is assumed  to be $n_e=5n_{cr}$ where $n_{cr}=1.38\times 10^{21}$~cm$^{-3}$, the laser beam power is 25~PW (the maximal field amplitude in mutual focus is $2600R_f$), the laser pulse duration is 30~fs and the pulse has a $sin^2$ envelope.

The simulation box was $264\times296\times296$ cells and $8.2\lambda\times9.1\lambda\times9.1\lambda$ along $x$, $y$ and $z$, where $\lambda=0.9$~$\mu$m. The time step was $T/64$, where $T\approx3$~fs. The time of simulations was $41T$. The larger simulation box and the longer time of simulation are necessary for the modeling of tightly focused laser pulses using the Total field – Scattered field technique \cite{taflove.fdtd.2005} by setting currents at the closed interface near the PML (perfectly matched layer) \cite{berenger.jcp.1994}. The PML ensures absorbing boundary condition for fields; particles and photons were also absorbed at the boundaries of simulation box. E-mode laser beams propagate along the $x$ axis and the largest component of the electric field is along the $z$ axis. The center of the seed sphere was located in the coordinate origin. Initially this sphere contains $10^7$ macroelectrons and hydrogen-like macroions, this quantity of macroparticles is equivalent to a statistical weight of particles of 1100. The threshold for resampling of particles and photons was $10^7$ macroparticles. We used the minimal sufficient spatial and time resolution for clearer demonstration of the capabilities of different event generators. Such a rough resolution can be used in practice during the optimization of various laser pulse or seed target parameters, which may require quite many simulations.

For these tests we used 2 two-CPU nodes of the Joint Supercomputer Center of RAS with Intel Xeon Gold 6248R. We run tests using 2 MPI processes with 96 threads per process.

In order to demonstrate agreement between simulations performed using different QED-event generators we show the time evolution of the quantity of electrons $N_e$ in the vicinity of the plane $y=0$ (-$0.03\lambda<y<0.03\lambda$)  (Fig.~\ref{fig:Real1}~(a)). At the beginning of interaction the initial target is deformed and destroyed by the leading edges of laser pulses. As a result, $N_e$ decreases between iterations $500<n_{it}<1000$. At this stage laser fields become relativistic but insufficient for fast QED processes. However, excessive sub-cycling within AEG increases the time required to compute the particle loop stage in comparison with the corresponding time in the cases of mAEG and FQED. For analysis this time was summed over each consecutive 100 iterations, we call the summed time $t_{100}$. FQED and mAEG show the same performance at this stage (Fig.~\ref{fig:Real1}~(b)). 

At the next stage laser fields become strong enough, a QED cascade is triggered and a rapid growth of $N_e$ occurs ($1000<n_{it}<1400$). Strong field amplitudes and a not very short time step allow FQED to reveal its remarkable performance at this stage: AEG is up to 6 times slower (mAEG is only slightly slower). 

When electron-positron plasma density becomes comparable with the relativistic critical density, the nonlinear self-consistent regime of laser interaction with extremely dense electron-positron plasma takes place ($1400<n_{it}<1600$). Field amplitudes sharply decrease inside dense plasma due to absorption and screening, so the speedup of particle loop computations with FQED and mAEG decreases (Fig.~\ref{fig:Real1}~(b)). Between FQED and mAEG the former shows slightly better performance.

The final stage is expansion of plasma due to decreasing of field intensity of incident pulses. Here performance becomes very similar with all three QED-event generators because $N_e$ strongly decreases (Fig.~\ref{fig:Real1}~(a)). We should note that in the case of mAEG $N_e$ shows deviation from that obtained with AEG and FQED. This discrepancy can be explained by several factors. First, $N_e(n_{it})$ is a random value, and, when the quantity of particles decreases, the dispersion of this random value increases. The reason of discrepancy of $N_e$ at $n_{it}\approx1000$ (at the end of the deformation and the destruction of the initial spherical target) is similar. Second, resampling leads to different ensembles of macroparticles in simulations with different QED-event generators. As a result, the absorption of macroparticles with different statistical weight at the boundaries of the simulation box additionally increase dispersion of $N_e$. 

As a result, a remarkable speedup of the particle loop stage with the FQED event generator during a third ($1000<n_{it}<1800$) of all iterations results in a decrease of the total particle loop time from 1952~s in the case of AEG to 786~s. This fact is demonstrated by the cumulative sum of $t_{100}$ in Fig.~\ref{fig:Real1}~(c).The mAEG event generator ensures performance (811~s for particle loop stage) very close to that in the case of FQED. The speedup for this problem is much less than that in the previous section and the reasons lie in the properties of interaction. Laser pulses have quite a short duration, so the maximal field amplitude is achieved for a relatively short period of time. The maximal fields are in quite a small region which particles escape relatively quickly. Also the dense generated electron-positron plasma decreases the field amplitude due to absorption and prevents penetration of strong fields inwards. Moreover, besides the particle loop stage there are other computational stages among which current deposition, communications and data output required approximately 900~s, 1100~s and 1100~s, respectively, independently on the type of QED-event generator. Thus the overall computational time demonstrates a less significant difference in performances: overall times were 6912~s, 5211~s and 5049~s with AEG, mAEG and FQED, respectively. Nevertheless, the newly developed QED-event generator can speed up simulations of such a problem, which is not very demanding for computation of QED processes, by 30\%, while the speedup of the particle loop stage can achieve the factor of 2.5.

\section{Conclusion}
In this paper, we proposed a new algorithm for taking into account QED effects in resource-intensive strong-field quantum electrodynamics numerical simulations. Unlike other methods, our scheme allows us to avoid excessive splitting of the time step, which leads to a substantial speed up of simulations. Our implementation is publicly available as part of the open-source hi-$\chi$ project designed as a Python-controlled toolbox for collaborative development. We plan to further optimize the code for better utilization of computational resources of modern CPUs.

\section{Authorship contribution statement}
\textbf{VV}: Conceptualization, software development, validation, verification, writing draft, editing.
\textbf{JM}: Conceptualization, validation, writing draft.
\textbf{AB}: Validation, verification, writing draft, editing.
\textbf{AM}: Validation, writing draft, editing.
\textbf{EE}: Conceptualization, supervision, validation, verification, writing draft, editing.
\textbf{IM}: Conceptualization, supervision, methodology, funding acquisition, writing draft, editing.

\section{Acknowledgements}
VV, AB, AM, EE, and IM acknowledge the use of computational resources provided by the Lobachevsky University and Joint Supercomputer Center of the Russian Academy of Sciences. 
VV and IM acknowledge the support of the Ministry of Science and Higher Education of the Russian Federation, project FSWR-2023-0034. AB and AM acknowledge the support of the Ministry of Science and Higher Education of the Russian Federation, project FFUF-2024-0030.

The authors would like to thank Arkady Gonoskov for useful discussions.

%\bibliography{literature}

\appendix
\section{Compton Rate Approximation}
\label{Appendix:ComptonRate}
The partial rate $\rd W/\rd\energy$ of nonlinear Compton scattering is given in equation~\ref{eq:photon-partial-rate}. In order to approximate the total rate function $W(\chi)$ we considered it as a sum of two functions:
\begin{equation}
    W(\chi)=C_1 (\chi)+C_2 (\chi),
\end{equation}
where $C_1$ and $C_2$ correspond to terms in (Eq.\ref{eq:photon-partial-rate}) containing the first and second synchrotron functions, respectively.
Each term was approximated independently by means of a local trend approximation with refinement using a polynomial factor in order to approximate the trend-free function:
\begin{equation}
    C_i(\chi)\approx Polynom(f(\chi)) \cdot  Trend(C_i (\chi)) 
\end{equation}

This approach factors out the main trend of the initial function, which often results in simpler forms of the approximating polynomial. The main trends of the Compton rate were chosen based on its asymptotic form:
\begin{equation}\label{eq:Wapprox}
W_\gamma(\chi) = \frac{\alpha}{\tau_C}\frac{1}{\gamma}
\left\lbrace\!\!\begin{array}{ccll}
\frac{5}{2\sqrt{3}}\chi &  \approx & 1.44\chi , & \text{if } \chi \ll 1 \\[\medskipamount]
\frac{14\Gamma(\frac{2}{3})}{9\sqrt[3]{3}} \chi^{2/3} &  \approx & 1.46\chi^{2/3} , & \text{if }  \chi \gg 1
\end{array}\right.,
\end{equation}
Basic information about the functions $C_1 (\chi)$ and $C_2 (\chi)$ is presented in Tables \ref{Table:C1} and \ref{Table:C2}, respectively. The exact form of the functions, including polynomial coefficients, is available in the open-source code hi-$\chi$ \cite{github_HiChi}.

\begin{table}[h!]
\centering
 \caption{Details of approximation of the $C_1(\chi)$ term of the Compton rate.}\label{Table:C1}
\begin{tabular}{ | m{3.4cm} | m{2.2cm}| m{1cm} | m{5.2cm} |  }
\hline
 $\chi$ interval  & $C_1(\chi)$ trend & $f(\chi)$ & Additional information about approximation of $C_1 (\chi)$ \\ 
 \hline 
 $\frac{2}{3\chi} < 10^{-12}$ & $\sqrt[3]{\frac{2}{3\chi}}$ & $\frac{2}{3\chi}$ & Degree $1$ polynomial \\ 
 \hline
 $10^{-12}\leqslant \frac{2}{3\chi} < 10^{-3}$ & $1$ & $\sqrt[6]{\frac{2}{3\chi}}$ & Degree $6$ polynomial \\ 
 \hline
 $10^{-3}\leqslant \frac{2}{3\chi} < 1$ & $1$ & $\sqrt[3]{\frac{2}{3\chi}}$ & Rational polynomial $\frac{P(f(\chi))}{Q(f(\chi))}$, degrees $6$ and $4$, respectively \\ 
 \hline
 $1\leqslant \frac{2}{3\chi} < 4$ & $\sqrt[3]{\left(\frac{2}{3\chi}\right)^2}$ & $\frac{2}{3\chi}$ & Degree $6$ polynomial \\ 
 \hline
 $4\leqslant \frac{2}{3\chi} < 100$ & $1$ & $\frac{2}{3\chi}$ & Degree $6$ polynomial \\ 
 \hline
 $\frac{2}{3\chi} \geqslant 100$ & $1$ & $\frac{3\chi}{2}$ & Degree $4$ polynomial \\ 
 \hline
 \end{tabular}
 \end{table}

\begin{table}[h!]
\centering
 \caption{Details of approximation of the $C_2(\chi)$ term of the Compton rate.}\label{Table:C2}
\begin{tabular}{ | m{3.4cm} | m{2.2cm}| m{1cm} | m{5.2cm} |  }
\hline
 $\chi$ interval  & $C_2(\chi)$ trend & $f(\chi)$ & Additional information about approximation of $C_2 (\chi)$ \\ 
 \hline 
 $\frac{2}{3\chi} < 10^{-3}$ & $1$ & $\sqrt[3]{\frac{2}{3\chi}}$ & Degree $3$ polynomial \\ 
 \hline
 $10^{-3}\leqslant \frac{2}{3\chi} < 4$ & $\sqrt[3]{\left(\frac{2}{3\chi}\right)^7}$;
 
 $\sqrt[3]{\left(\frac{2}{3\chi}\right)^{11}}$;
 
 $\sqrt[3]{\frac{2}{3\chi}}$;
 
 $\sqrt[3]{\left(\frac{2}{3\chi}\right)^5}$;
 
 $\frac{2}{3\chi}$; 
 
 $\left(\frac{2}{3\chi}\right)^2$ & $\left(\frac{2}{3\chi}\right)^2$ & $C_2(\chi)$ considered as a sum of $6$ functions with individual trends, these functions are approximated independently. For each term the degree of the polynomial is $10$. \\
 \hline
 $4\leqslant \frac{2}{3\chi} < 100$ & $1$ & $\sqrt[4]{\frac{2}{3\chi}}$ & Rational polynomial $\frac{P(f(\chi))}{Q(f(\chi))}$, degrees $10$ and $2$, respectively \\ 
 \hline
 $\frac{2}{3\chi} \geqslant 100$ & $1$ & $\frac{3\chi}{2}$ & Degree $7$ polynomial \\ 
 \hline
 \end{tabular}
 \end{table}

\section{Compton inverse CDF Approximation}
\label{Appendix:ComptonCDF}

As described earlier in Sec.\ref{sec:comptonCDF}, in the main domain linear interpolation based on precomputed and tabulated values on the mesh is employed.

At small values of $\energy$, the cumulative distribution function scales as
\begin{equation}
    \cdf_\gamma(\chi, \energy) = \frac{1}{W(\chi)}\frac{2 \cdot 3^{2/3}}{\Gamma(\frac{1}{3})} \energy^{1/3} \chi^{2/3}.
\end{equation}

Therefore, the Compton inverse CDF function at the lower bound of $r_2$ ($r_2<0.05$) is approximated by the following asymptotic form:
\begin{equation}
    \cdf_\gamma^{-1}(\chi, r_2) = D(\chi) \cdot \sqrt[3]{r_2}
\end{equation}
The coefficient $D(\chi)$ is calculated from the conditions of the Compton inverse CDF function continuity at the closest mesh nodes.

\section{Breit-Wheeler Rate Approximation}
\label{Appendix:BWRate}
In order to approximate the function $W_p(\chi_\gamma)$ we consider it as a sum of two functions:
\begin{equation}
    W_p(\chi_\gamma)=B_1 (\chi_\gamma)+B_2 (\chi_\gamma),   
\end{equation}
corresponding to terms in (\ref{eq:pair-partial-rate}) containing the first and second synchrotron functions, respectively.
Each term is again approximated independently using the same approach as in \ref{Appendix:ComptonRate}:
\begin{equation}\label{eq:C1}
    B_i(\chi_\gamma)\approx Polynom(f(\chi_\gamma)) \cdot Trend(B_i (\chi_\gamma))
\end{equation}
The main trends of the Breit-Wheeler rate were chosen based on its asymptotic form:
\begin{equation}\label{eq:Wpairs}
W_p(\chi_\gamma) = \frac{\alpha}{\tau_C} \frac{mc^2}{\hbar\omega}
\left\lbrace\!\!\begin{array}{ccll}
\frac{3\sqrt{3}}{16\sqrt{2}} \chi_\gamma \e^{- 8/3\chi_\gamma}  & \approx & 0.23\chi_\gamma \e^{- 8/3\chi_\gamma}, & \text{if } \chi_\gamma \ll 1 \\[\medskipamount]
\frac{20\pi^2}{7 \sqrt[3]{3} \Gamma(\frac{1}{3})^4} \chi_\gamma^{2/3}  & \approx & 0.38\chi_\gamma^{2/3}, & \text{if }  \chi_\gamma \gg 1
\end{array}\right.,
\end{equation}
Basic information about the functions $B_1(\chi_\gamma)$ and $B_2(\chi_\gamma)$ is presented in Tables \ref{Table:B1} and \ref{Table:B2}, respectively.

\begin{table}[h!]
\centering
 \caption{Details of approximation of the $B_1(\chi_\gamma)$ term of the Breit-Wheeler rate.}\label{Table:B1}
\begin{tabular}{ | m{3.4cm} | m{2.2cm}| m{1cm} | m{5.2cm} | }
\hline
 $\chi_\gamma$ interval  & $B_1(\chi_\gamma)$ trend & $f(\chi_\gamma)$ & Additional information about approximation of $B_1 (\chi_\gamma)$ \\ 
 \hline 
 $\frac{8}{3\chi_\gamma} < 10^{-6}$ & $1$ & $\sqrt[3]{\frac{8}{3\chi_\gamma}}$ & Degree $4$ polynomial \\
 \hline
 $10^{-6}\leqslant \frac{8}{3\chi_\gamma} < 10^{-1}$ & $1$ & $\sqrt[3]{\frac{8}{3\chi_\gamma}}$ & Degree $8$ polynomial \\ 
 \hline
 $10^{-1}\leqslant \frac{8}{3\chi_\gamma} < 1$ & $e^{-\frac{8}{3\chi_\gamma}}$ & $\sqrt[3]{\frac{8}{3\chi_\gamma}}$ & Degree $7$ polynomial \\ 
 \hline
 $1\leqslant \frac{8}{3\chi_\gamma} < 4$ & $e^{-\frac{8}{3\chi_\gamma}}$ & $\sqrt[3]{\frac{8}{3\chi_\gamma}}$ & Degree $7$ polynomial \\ 
 \hline
 $4\leqslant \frac{8}{3\chi_\gamma} < 200$ & $e^{-\frac{8}{3\chi_\gamma}}$ & $\frac{3\chi_\gamma}{8}$ & Degree $7$ polynomial \\ 
 \hline
 $\frac{8}{3\chi_\gamma} \geqslant 200$ & $e^{-\frac{8}{3\chi_\gamma}}$ & $1$ & Without polynomial \\ 
 \hline
 \end{tabular}
 \end{table}

\begin{table}[h!]
\centering
 \caption{Details of approximation of the $B_2(\chi_\gamma)$ term of the Breit-Wheeler rate.}\label{Table:B2}
\begin{tabular}{ | m{3.4cm} | m{2.2cm}| m{1cm} | m{5.2cm} |  }
\hline
 $\chi_\gamma$ interval  & $B_2(\chi_\gamma)$ trend & $f(\chi_\gamma)$ & Additional information about approximation of $B_2 (\chi_\gamma)$ \\ 
 \hline 
 $\frac{2}{3\chi_\gamma} < 2^{-6}$ & ${\left(\frac{3\chi_\gamma}{2}\right)}^4$ & $\sqrt[3]{\frac{3\chi_\gamma}{2}}$ & Degree $5$ polynomial \\ 
 \hline
 $2^{-6}\leqslant \frac{2}{3\chi_\gamma} < 2^{-3}$ & $e^{-\frac{8}{3\chi_\gamma}}$ & $\sqrt[3]{\frac{2}{3\chi_\gamma}}$; $\sqrt[6]{\frac{3\chi_\gamma}{2}}$ & \multirow{3}{5.2cm}{
 The coefficient $Polynom(f(\chi_\gamma))$ in \ref{eq:C1} is computed as the sum of two polynomials of degree 7
 } \\
 \cline{1-3}
 $2^{-3}\leqslant \frac{2}{3\chi_\gamma} < 2^{-1}$ & $e^{-\frac{8}{3\chi_\gamma}}$ & $\sqrt[3]{\frac{2}{3\chi_\gamma}}$; $\sqrt[4]{\frac{3\chi_\gamma}{2}}$ & \\ 
 \cline{1-3}
 $2^{-1}\leqslant \frac{2}{3\chi_\gamma} < 1$ & $e^{-\frac{8}{3\chi_\gamma}}$ & $\sqrt[6]{\frac{2}{3\chi_\gamma}}$; $\sqrt[4]{\frac{3\chi_\gamma}{2}}$ & \\ 
 \hline
 $1\leqslant \frac{2}{3\chi_\gamma} < 8$ & $e^{-\frac{8}{3\chi_\gamma}}$ & $\sqrt[6]{\frac{2}{3\chi_\gamma}}$ & Degree $7$ polynomial \\ 
 \hline
 $8\leqslant \frac{2}{3\chi_\gamma} < 50$ & $e^{-\frac{8}{3\chi_\gamma}}$ & $\frac{3\chi_\gamma}{2}$ & Degree $7$ polynomial \\ 
 \hline
 $\frac{2}{3\chi} \geqslant 50$ & $e^{-\frac{8}{3\chi_\gamma}}$ & $\frac{3\chi_\gamma}{2}$ & Degree $7$ polynomial \\ 
 \hline
 \end{tabular}
 \end{table}

\section{Breit-Wheeler Inverse CDF Approximation}
\label{Appendix:BWCDF}

As described above in Sec.\ref{sec:BWCDF}, linear interpolation based on precomputed and tabulated values on the mesh is employed in the main domain.

At small values of $\energy_e$, the cumulative distribution function scales as follows:
\begin{equation}
    \cdf_p(\chi_\gamma, \energy_e) = \frac{1}{W(\chi_\gamma)}\frac{3}{8\sqrt{\pi}} (\chi_\gamma\energy_e)^{3/2} \exp(-\frac{2}{3\chi_\gamma\energy_e})
\end{equation}

Therefore, the Breit-Wheeler inverse CDF function at the lower bound of $r_2$ ($r_2<10^{-5}$) is approximated by the following asymptotic form:

 \begin{equation}
    \cdf_p^{-1}(\chi_\gamma, r_2) = L(\chi_\gamma) \cdot \frac{\chi_\gamma}{LambertW\left[M(\chi_\gamma) \cdot \left( \frac{\chi_\gamma}{r_2} \right)^{\frac{2}{3}}\right]},
\end{equation}
where the $LambertW$ function \cite{corless1996lambertw} is employed. The coefficients $L$ and $M$ are calculated from the conditions of function continuity on the closest mesh nodes.
\end{document}